\documentclass[preprint,aps,12pt,showpacs,nofootinbib,tightenlines]{revtex4-1}
\usepackage{amsmath}
\usepackage{amssymb}
\usepackage{epsfig}
\usepackage{graphicx}
\begin{document}

\newcommand{\beq}{\begin{eqnarray}}
\newcommand{\eeq}{\end{eqnarray}}
\newcommand{\non}{\nonumber\\ }

\newcommand{\jpsi}{J/\Psi}

\newcommand{\pka}{\phi_{K}^A}
\newcommand{\pkp}{\phi_K^P}
\newcommand{\pkt}{\phi_K^T}

\newcommand{\pea}{\phi_{\eta}^A}
\newcommand{\pep}{\phi_{\eta}^P}
\newcommand{\pet}{\phi_{\eta}^T}
\newcommand{\peqa}{\phi_{\eta_q}^A}
\newcommand{\peqp}{\phi_{\eta_q}^P}
\newcommand{\peqt}{\phi_{\eta_q}^T}

\newcommand{\pesa}{\phi_{\eta_s}^A}
\newcommand{\pesp}{\phi_{\eta_s}^P}
\newcommand{\pest}{\phi_{\eta_s}^T}

\newcommand{\pecv}{\phi_{\eta_c}^v}
\newcommand{\pecs}{\phi_{\eta_c}^s}

\newcommand{\pepa}{\phi_{\eta'}^A}
\newcommand{\pepp}{\phi_{\eta'}^P}
\newcommand{\pept}{\phi_{\eta'}^T}

\newcommand{\pksa}{\phi_{K^*}}
\newcommand{\pksp}{\phi_{K^*}^s}
\newcommand{\pkst}{\phi_{K^*}^t}

\newcommand{\fb}{f_B }
\newcommand{\fk}{f_K }
\newcommand{\fe}{f_{\eta} }
\newcommand{\fep}{f_{\eta'} }
\newcommand{\rks}{r_{K^*} }
\newcommand{\rk}{r_K }
\newcommand{\re}{r_{\eta} }
\newcommand{\rep}{r_{\eta'} }
\newcommand{\mb}{m_B }
\newcommand{\mw}{m_W }
\newcommand{\im}{{\rm Im} }

\newcommand{\kks}{K^{(*)}}
\newcommand{\etar}{\eta^{\prime}}
\newcommand{\acp}{{\cal A}_{CP}}
\newcommand{\etap}{\eta^{(\prime)} }
\newcommand{\pb}{\phi_B}
\newcommand{\pks}{\phi_{K^*}}

\newcommand{\xeba}{\bar{x}_2}
\newcommand{\xsba}{\bar{x}_3}
\newcommand{\res}{r_{\eta_s}}
\newcommand{\red}{r_{\eta_q}}
\newcommand{\peas}{\phi^A_{\eta_s}}
\newcommand{\peps}{\phi^P_{\eta_s}}
\newcommand{\pets}{\phi^T_{\eta_s}}
\newcommand{\pead}{\phi^A_{\eta_q}}
\newcommand{\pepd}{\phi^P_{\eta_q}}
\newcommand{\petd}{\phi^T_{\eta_q}}

\newcommand{\pvsl}{ p \hspace{-2.0truemm}/_{K^*} }
\newcommand{\esl}{ \epsilon \hspace{-2.1truemm}/ }
\newcommand{\psl}{ P \hspace{-2.4truemm}/ }
\newcommand{\nsl}{ n \hspace{-2.2truemm}/ }
\newcommand{\vsl}{ v \hspace{-2.2truemm}/ }
\newcommand{\epsl}{\epsilon \hspace{-1.8truemm}/\,  }
\newcommand{\bfkk}{{\bf k} }
\newcommand{\calm}{ {\cal M} }
\newcommand{\calh}{ {\cal H} }

\def \cpc{ Chin. Phys. C }
\def \epjc{ Eur. Phys. J. C }
\def \jpg{  J. Phys. G }
\def \npb{  Nucl. Phys. B }
\def \plb{  Phys. Lett. B }
\def \pr{  Phys. Rep. }
\def \prd{  Phys. Rev. D }
\def \prl{  Phys. Rev. Lett.  }
\def \zpc{  Z. Phys. C  }
\def \jhep{ J. High Energy Phys.  }

\title{Anatomy of $B \to K \eta^{(\prime)}$ decays in different mixing schemes and
effects of next-to-leading order contributions in the perturbative QCD approach}
\author{Ying-Ying Fan, Wen-Fei Wang, Shan Cheng, and Zhen-Jun Xiao
\footnote{Electronic address: xiaozhenjun@njnu.edu.cn},}
\affiliation{Department of Physics and Institute of Theoretical Physics,
Nanjing Normal University, Nanjing, Jiangsu 210023, P.R.China }
\date{\today}
\begin{abstract}
In this paper we perform a comprehensive study of the four $B \to K
\etap$ decays in the perturbative QCD (pQCD) factorization approach.
We  calculate the CP-averaged branching ratios and CP-violating
asymmetries of $B \to K\etap$ decays in the ordinary $\eta$-$\etar$,
the $\eta$-$\etar$-$G$, and the
$\eta$-$\etar$-$G$-$\eta_c$ mixing schemes. Besides the
full leading order (LO) contributions, all currently known
next-to-leading order (NLO) contributions to $B \to K \etap$ decays
in the pQCD approach are taken into account. From our calculations
and phenomenological analysis, we find the following. (a) The NLO
contributions in general can provide significant enhancements to the
LO pQCD predictions for the decay rates of the two $B \to K \etar$
decays, around $70\%-89\%$ in magnitude, but result in relatively
small changes to $Br(B \to K \eta)$.
(b) Although the NLO pQCD
predictions in all three considered mixing schemes agree well with
the data within one standard deviation, those pQCD predictions in
the $\eta$-$\etar$-$G$ mixing scheme provide the best interpretation
for the measured pattern of $Br(B \to K \etap)$: $Br(B^0 \to K^0
\eta) \approx 1.13\times 10^{-6}$, $Br(B^0 \to K^0 \etar) \approx
66.5 \times 10^{-6}$, $Br(B^\pm \to K^\pm \eta) \approx 2.36\times
10^{-6}$, and $Br(B^\pm \to K^\pm \etar) \approx 67.3 \times
10^{-6}$, which agree perfectly with the measured values.
(c) The NLO pQCD predictions for the CP-violating asymmetries for the four
considered decays are also in good consistent with the data.
(d) The newly known NLO contribution to the relevant form factors
$\calm_{FF}$ can produce about a $20\%$ enhancement to the branching
ratios $Br(B \to K \etar)$, which plays an important role in closing
the gap between the pQCD predictions and the relevant data.
(e) The general expectations about the relative strength of the LO and NLO
contributions from different sources are examined and confirmed by
explicit numerical calculations.
\end{abstract}

\vspace{1cm}


\maketitle

\section{Introduction}

Since the first observation of unexpectedly large branching ratios
for $B \to K \etar$ decays was reported by CLEO in 1997
\cite{cleo97}, the four $B\to K\etap$ decays have been a ``hot"
topic for a long time. Although many physicists have made great
efforts to explain the pattern of very large branching ratios $Br(B
\to K \etar)$ and very small branching ratios $Br(B \to K \eta)$
\cite{kagan,xiao99,yy01,kls01,lu2001,bn651,kou02,li2003}, it is
still difficult to understand such a pattern in the framework of the
standard model(SM).

On the experimental side, the branching ratios of the four $B\to K \etap$ decays have been
measured with high precision \cite{hfag,pdg2012},
\beq
Br(B^0 \to K^0 \eta)&=& (1.23^{+0.27}_{-0.24}) \times 10^{-6}, \non
Br(B^0 \to K^0 \etar)&=& (66.1\pm 3.1) \times 10^{-6}, \non
Br(B^\pm \to K^\pm \eta)&=& (2.36^{+0.22}_{-0.21}) \times 10^{-6}, \non
Br(B^\pm \to K^\pm \etar)&=& (71.1\pm 2.6) \times 10^{-6},
 \label{eq:br-k}
\eeq
For the relevant CP-violating asymmetries, the currently known experimental
measurements are \cite{hfag,pdg2012}
\beq
\acp^{dir}(B^0 \to K^0 \etar)&=& (1 \pm 9)\%, \non
\acp^{mix}(B^0 \to K^0 \etar)&=& (64 \pm 11)\%, \non
\acp^{dir}(B^\pm \to K^\pm \eta)&=& (-37 \pm 8)\%, \non
\acp^{dir}(B^\pm \to K^\pm \etar)&=& (1.3 ^{+1.6}_{-1.7})\%,
\label{eq:cp-k}
\eeq

On the theory side,  these decays were calculated recently in
Ref.~\cite{zjx2008} by employing the perturbative QCD (pQCD)
factorization approach\cite{li2003} and using the ordinary Feldmann-Kroll-Stech
(FKS) $\eta-\etar$ mixing scheme \cite{fks98}, with the inclusion of
the partial next-to-leading order (NLO) contributions, i.e.,
the QCD vertex corrections, the
quark-loops, and the chromomagnetic penguins $O_{8g}$. In
Ref.~\cite{zjx2008} the authors found that the NLO contributions can
provide a $70\%$ enhancement to the leading order (LO) pQCD predictions for $Br(B
\to K \etar)$, but also produce a $30\%$ reduction to the LO pQCD predictions for
$Br(B \to K \eta)$ \cite{zjx2008}; numerically,  $Br(B^0 \to K^0 \eta)
\approx 2.1 \times 10^{-6}$, $Br(B^0 \to K^0 \etar) \approx 50.3
\times 10^{-6}$, $Br(B^+ \to K^+ \eta) \approx 3.2 \times 10^{-6}$,
and $Br(B^\pm \to K^\pm \etar) \approx 51.0 \times 10^{-6}$.
Although the differences between the pQCD predictions and the data
were effectively decreased by the inclusion of the partial NLO
contributions, the central values of the pQCD predictions for $Br(B
\to K \etar)$ are still lower than the data by about $30\%$. As for
the CP-violating asymmetries, the pQCD predictions in Ref.~\cite{zjx2008}
already agree well with the data.

Very recently, three new advances in the studies of the two-body charmless
hadronic $B\to M_2 M_3$ decays (here $M_i$ stands for the light mesons, such as
$\pi, K, \etap, etc.$) in the pQCD factorization approach have been made:
\begin{itemize}
\item[(i)]
In Ref.~\cite{prd85074004}, Li et al. calculated the NLO contributions to the form factors of $B \to \pi$ transitions in
the pQCD approach and found that the NLO part can provide a roughly
$20\%$ enhancement to the LO one.
The enhanced form factors may then lead to a larger branching ratio for $B \to K \etar$
decays. The still missing NLO parts in the pQCD approach are the
$O(\alpha_s^2)$ contributions from nonfactorizable
spectator diagrams and the annihilation diagrams, which are most likely
small according to general arguments.

\item[(ii)]
In Ref.~\cite{prd86011501}, the authors studied and provided a successful
pQCD interpretation for the Belle measurements of
$B_d/B_s \to \jpsi  \etap$, i.e., $R_{q}^{exp}=Br( B_{q}\to \jpsi \etar)/Br(B_{q}\to \jpsi \eta) < 1 $ with $q=(d,s)$,
by using the $\eta$-$\eta'$-$G$ mixing
formalism\cite{eepg}, where $G$ represents the pseudoscalar glueball.
This result encourages us to check the possible effects of the
pseudoscalar $G$ on $B \to K \etap$ decays, although such contributions may be small as generally
expected \cite{li2008}.

\item[(iii)]
In Ref.~\cite{eepgec}, the authors studied the  $\eta$-$\eta'$-$G$-$\eta_c$
mixing scheme, obtained constraints
on the mixing angle $\phi_Q$ ($\phi_Q\sim 11^\circ$)
between $G$ and $\eta_c$ by fitting to the observed
$\eta_c$ decay widths and other relevant data, and found that the $\eta_c$
mixing  can enhance the pQCD predictions for
$Br(B \to K \etar)$ by $18\%$, but does not alter those for $Br(B \to K \eta)$.
In Ref.~\cite{eepgec}, the authors superposed the contributions
from $B \to K \eta_c$ due to the $\eta_c$ mixing onto
the partial NLO pQCD predictions as given in Ref.~\cite{zjx2008}
directly. They did not consider the effects
of the newly known NLO contributions to the corresponding
form factors $F_0^{B\to K}(0)$ and $F_0^{B\to \etap}(0)$.

\end{itemize}

Motivated by the above new advances\cite{prd85074004,prd86011501,eepgec}, we think that it is
time for us to make a comprehensive study of the four $B \to K \etap$ decays
in the pQCD factorization approach. We will focus on the following points:
\begin{itemize}
\item[(i)]
Besides those NLO contributions already considered in Ref.~\cite{zjx2008},
we here will firstly extend the calculation of the NLO
part of the form factors for the $B \to \pi$ transition \cite{prd85074004}
to the cases for the similar $B \to K$ and $B \to (\eta_q,\eta_s)$ transitions,
and then take these newly known form factors at the NLO level into the calculations
for the branching ratios and CP-violating asymmetries of $B \to K \etap$
decays, to check its effects on the corresponding pQCD predictions.

\item[(ii)]
Besides the ordinary  FKS $\eta$-$\etar$ mixing scheme \cite{fks98}, we will also calculate these four decays in the $\eta$-$\eta'$-$G$ \cite{eepg} mixing scheme  and the $\eta$-$\eta'$-$G$-$\eta_c$ mixing scheme \cite{eepgec}, respectively.
We want to check the effects of the possible "pseudoscalar glueball"
and $\eta_c$ component  of the $\etap$ meson on the pQCD predictions.

\item[(iii)]
We will numerically calculate the individual decay amplitudes
$\calm^{a+b}$ [ the emission diagrams Figs.1(a) and 1(b)],
$\calm^{c+d}$ [the spectator diagrams Figs.1(c) and 1(d)] and $\calm^{anni}$
[annihilation diagrams Figs.1(e)-1(h) ]
and compare the relative strength of the contributions from different
sets of the Feynman  diagrams at the leading order,
or from the different sources at the next-to-leading  order,
such as $\calm_{VC}$ (i.e. NLO vertex corrections
Figs.3(a)-3(d) ), $\calm_{ql}$ [ i.e., NLO quark-loops
Figs.3(e) and 3(f)], and $\calm_{mp}$ [ i.e., NLO chromomagnetic
penguins Figs.3(g) and 3(h)]. We try to find the source
of the dominant contribution, in order to estimate the
possible strength of the two still missing NLO contributions in the pQCD approach.

\end{itemize}

The paper is organized as follows. In Sec.~II, we give a brief discussion of
the pQCD factorization approach
and the three different kinds of mixing schemes: $\eta$-$\etar$,
$\eta$-$\etar$-$G$, and $\eta$-$\etar$-$G$-$\eta_c$ mixing schemes.
In Sec.~III, we make the analytic calculations of the relevant
Feynman diagrams and present the various decay amplitudes for the
studied decay modes in leading order. In Sec.~IV,
all currently known NLO contributions in the pQCD approach are investigated.
In Sec.~V, we will show the numerical results for the pQCD predictions for
the branching ratios and CP-violating asymmetries of the considered decay modes,
and calculate and compare the relative strength of the LO and NLO
contributions from different sets of the Feynman  diagrams or from different sources.
The summary and some discussions are included in the final section.


\section{ Theoretical framework}\label{sec:f-work}

As is well known, the pQCD factorization approach has been widely
used in studies for the two-body
charmless hadronic $B/B_s/B_c \to M_2 M_3$ decays
(here $M_i$ stands for the light pseudoscalar meson $P$,
the vector meson $V$, the scalar meson $S$, etc.)
\cite{kls01,kou02,li2003,zjx2008,prd85074004,zjx2006b,alibs,nlo05,prd63,liux10,liubc,xiao2012}.
Some pQCD predictions-for example, the large CP-violating asymmetries $\acp^{dir}(B^0\to \pi^+\pi^-)\approx (30\pm10)\%$
in Ref.~\cite{lu2001} and the large branching ratio $Br(B_s\to \pi^+\pi^-) \approx 5\times 10^{-6}$
in Refs.~\cite{liy2004,alibs}, -have been confirmed by experiments\cite{xiao2012}.
We here focus on the study of the four $B \to K \etap$ decays. For more details of the formalism of the pQCD factorization
approach itself, one can see, for example, Refs.~\cite{li2003,zjx2008,nlo05,li2011}.

\subsection{Outline of $B \to K \etap$ decays in the pQCD approach}\label{sec:2-1}

In the B rest frame,  we assume that the light final-state meson
$M_2$ and $M_3$ is moving along the direction of $n=(1,0,
{\bf{0}}_T)$ and $v=(0,1,{\bf{0}}_T)$, respectively. We use $x_i$ to
denote the momentum fraction of the antiquark in each meson,
and $k_{i\perp}$ the corresponding transverse momentum. Using the
light-cone coordinates the $B$-meson momentum $P_B$ and the two
final-state meson's momenta $P_2$ and $P_3$ (for $M_2$ and $M_3$
respectively) can be written as
\beq
P_B = \frac{M_B}{\sqrt{2}} (1,1,{\bf 0}_{\rm T}), \quad
P_2 =\frac{M_B}{\sqrt{2}}(1-r_3^2,r^2_2,{\bf 0}_{\rm T}), \quad
P_3 =\frac{M_B}{\sqrt{2}} (r_3^2,1-r^2_2,{\bf 0}_{\rm T}),
\eeq
where $r_i=m_i/M_B$ with $m_i$ being the mass of meson $M_i$. If we choose
\beq
k_1 &=& \frac{m_B}{\sqrt{2}}\left (x_1,0,{\bf k}_{\rm 1T}\right), \quad
k_2 =\frac{m_B}{\sqrt{2}}\left (x_2(1-r_3^2),x_2 r^2_2,{\bf k}_{\rm 2T} \right ),\non
k_3&=& \frac{m_B}{\sqrt{2}}\left (x_3
r_3^2,x_3(1-r_2^2),{\bf k}_{\rm 3T} \right),
\eeq
The integration over the small components $k_1^-$, $k_2^-$, and $k_3^+$ will lead
conceptually to the decay amplitudes,
\beq
{\cal A}(B \to M_2M_3 ) &\sim &\int\!\! d x_1 d x_2 d x_3 b_1 d b_1 b_2 d b_2 b_3 d b_3 \non &&
\cdot \mathrm{Tr} \left [ C(t) \Phi_B(x_1,b_1) \Phi_{M_2}(x_2,b_2)
\Phi_{M_3}(x_3, b_3) H(x_i, b_i, t) S_t(x_i)\, e^{-S(t)} \right ],
\quad \label{eq:a2}
\eeq
where $b_i$ is the conjugate space coordinate of $k_{iT}$.
In the above equation, $C(t)$ is the Wilson
coefficient evaluated at scale $t$, which includes the large logarithms ($\ln m_W/t$)
coming from QCD radiative corrections to four-quark operators.
The functions $\Phi_B(x_1,b_1)$, $\Phi_{M_2}(x_2,b_2)$, and
$\Phi_{M_3}(x_3,b_3)$ are the wave functions of the initial B meson and the
final-state meson $M_2$ and $M_3$ respectively. These wave functions describe
the hadronization of the quark and antiquark in the meson $B$ and $M_{2,3}$.
The ``hard kernel" $H(k_1,k_2,k_3,t)$ describes the four-quark operator and the
spectator quark connected by  a hard gluon whose $q^2$ is in the order
of $\bar{\Lambda} M_B$, and includes the $\mathcal{O}(\sqrt{\bar{\Lambda} M_B})$ hard
dynamics.

The jet function $S_t(x_i)$ in Eq.(\ref{eq:a2})$-$as given explicitly
in Eq.(\ref{eq:stxi}) of Appendix B$-$
is one of the two kinds of Sudakov form factors relevant for the B decays considered,
which come from the threshold resummation over the large double logarithms ($\ln^2 x_i$)
in the end-point region. This jet function $S_t(x_i)$ vanishes as $x_i\to 0,1$,
and smears the end-point singularities on $x_i$ for meson distribution amplitudes.

Similarly, the inclusion of $k_{\rm T}$ regulates the end-point singularities.
The large double logarithms $\alpha_s \ln^2 k_{\rm T}$ should also be organized
to all orders, leading to a $k_{\rm T}$ resummation \cite{collins81}.
The resultant Sudakov form factors $S_B(t), S_2(t)$ and $S_3(t)$ for the B meson
and two final-state mesons $M_{2,3}$$-$ whose explicit expressions can be found in
Eq.(B9) of Appendix B $-$ keep the magnitude of $k_{\rm T}^2$ to be at roughly
${\cal O}(\overline{\Lambda} m_B)$ by suppressing the region
with $k_{\rm T}^2\sim {\cal O}(\overline{\Lambda}^2)$.
The exponential function $e^{-S(t)}$ in Eq.(\ref{eq:a2})$-$
where $s(t)=S_B(t)+S_2(t)+S_3(t)$
or $S(t)=S_B(t)+S_i(t)$ with $i=2$ or $3$ as shown in Eq.(B8) of Appendix B $-$
is also called the Sudakov form factor, which effectively suppresses the soft dynamics
at the end-point region \cite{li2003}. Of course, more studies are
required to check the actual suppression effect on possible nonperturbative
contributions due to the introduction of the Sudakov form factors. Some theoretical
errors may also be produced due to the uncertainties of $S_t(x_i)$ and
$e^{-S(t)}$.

For the studied $B \to K \etap$ decays, the corresponding weak
effective Hamiltonian can be written as \cite{buras96}
\beq
{\cal H}_{eff} &=& \frac{G_{F}}{\sqrt{2}}     \Bigg\{ V_{ub} V_{uq}^{\ast} \Big[
     C_{1}({\mu}) O^{u}_{1}({\mu})  +  C_{2}({\mu}) O^{u}_{2}({\mu})\Big]
  -V_{tb} V_{tq}^{\ast} \Big[{\sum\limits_{i=3}^{10}} C_{i}({\mu}) O_{i}({\mu})
  \Big ] \Bigg\} + \mbox{H.c.} ,
 \label{eq:heff}
\eeq where $q=d,s$, $G_{F}=1.166 39\times 10^{-5} GeV^{-2}$ is the
Fermi constant. The $O_{i}$ ($i=1,...,10$) are the local four-quark operators,
\beq
O^{u}_{1}&=& ({\bar{u}}_{\alpha}b_{\beta} )_{V-A}({\bar{q}}_{\beta} u_{\alpha})_{V-A},
    \quad \quad O^{u}_{2}=({\bar{u}}_{\alpha}b_{\alpha})_{V-A}({\bar{q}}_{\beta} u_{\beta} )_{V-A},\\
O_{3}&=&({\bar{q}}_{\alpha}b_{\alpha})_{V-A}\sum\limits_{q^{\prime}}
           ({\bar{q}}^{\prime}_{\beta} q^{\prime}_{\beta} )_{V-A},
    \quad
    O_{4}=({\bar{q}}_{\beta} b_{\alpha})_{V-A}\sum\limits_{q^{\prime}}
           ({\bar{q}}^{\prime}_{\alpha}q^{\prime}_{\beta} )_{V-A},\non
O_{5}&=&({\bar{q}}_{\alpha}b_{\alpha})_{V-A}\sum\limits_{q^{\prime}}
           ({\bar{q}}^{\prime}_{\beta} q^{\prime}_{\beta} )_{V+A},
    \quad
O_{6}=({\bar{q}}_{\beta} b_{\alpha})_{V-A}\sum\limits_{q^{\prime}}
           ({\bar{q}}^{\prime}_{\alpha}q^{\prime}_{\beta} )_{V+A},\\
O_{7}&=&\frac{3}{2}({\bar{q}}_{\alpha}b_{\alpha})_{V-A} \sum\limits_{q^{\prime}}e_{q^{\prime}}
           ({\bar{q}}^{\prime}_{\beta} q^{\prime}_{\beta} )_{V+A},
\quad
O_{8}=\frac{3}{2}({\bar{q}}_{\beta} b_{\alpha})_{V-A}      \sum\limits_{q^{\prime}}e_{q^{\prime}}
           ({\bar{q}}^{\prime}_{\alpha}q^{\prime}_{\beta} )_{V+A},\non
O_{9}&=&\frac{3}{2}({\bar{q}}_{\alpha}b_{\alpha})_{V-A} \sum\limits_{q^{\prime}}e_{q^{\prime}}
           ({\bar{q}}^{\prime}_{\beta} q^{\prime}_{\beta} )_{V-A},\quad
O_{10}=\frac{3}{2}({\bar{q}}_{\beta} b_{\alpha})_{V-A}           \sum\limits_{q^{\prime}}e_{q^{\prime}}
           ({\bar{q}}^{\prime}_{\alpha}q^{\prime}_{\beta} )_{V-A},
\eeq
where $\alpha$ and $\beta$ are the color indices and $q^\prime$ are
the active quarks at the scale $m_b$, i.e.  $q^\prime=(u,d,s,c,b)$.
The left-handed current is defined as $({\bar{q}}_{\alpha}
q_{\beta} )_{V-A}= {\bar{q}}_{\alpha} \gamma_\nu (1-\gamma_5) q_{\beta}  $
and the right-handed current
$({\bar{q}}_{\alpha} q_{\beta} )_{V+A}={\bar{q}}_{\alpha} \gamma_\nu (1+\gamma_5)q_{\beta}  $.

In this paper, we will calculate $B \to K \etap$
decays in the pQCD approach with the inclusion of all known
NLO contributions, and focus on the effects of the
newly known NLO contributions to the form factors
$F_0^{B\to K}(0)$ and $F_0^{B\to \etap}(0)$ \cite{prd85074004}.

\subsection{ Different mixing schemes}\label{sec:2-2}

Both $\eta$ and $\etar$ mesons $-$ according to currently available studies
\cite{fks98,eepg,eepgec,en2007} $-$ may contain a small gluonic
component ( an $\eta_c$ or even a $\pi^0$ component)
through mixing.

In order to check the mixing-scheme dependence of the pQCD predictions
for the physical observables of the considered decays
we will calculate the $B\to K \etap$ decays in the following three typical
mixing schemes(MS):
\begin{enumerate}
\item[(i)]
MS-1: The FKS scheme \cite{fks98} of $\eta$-$\etar$ mixing in the
quark-flavor basis \footnote{In the octet-singlet basis, one assumes
$\eta_1=(\bar{u}u+\bar{d}d+\bar{s}s)/\sqrt{3}$ and $\eta_8=(\bar{u}u+\bar{d}d-2\bar{s}s)/\sqrt{6}$. }:
$\eta_q= (u\bar u +d\bar d)/\sqrt{2}$ and $\eta_s=s\bar{s}$.

\item[(ii)]
MS-2: The $\eta$-$\etar$-$G$ mixing scheme as defined in Ref.~\cite{eepg}.

\item[(iii)]
MS-3: The $\eta$-$\etar$-$G$-$\eta_{c}$ mixing scheme as defined in Ref.~\cite{eepgec}.

\end{enumerate}

Firstly, in the FKS $\eta$-$\etar$ mixing scheme in the quark-flavor basis,
the physical $\eta$ and $\etar$ can be written as
\beq
\left(\begin{array}{c} \eta \\ \eta^{\prime}\end{array} \right)
= U(\phi) \left(\begin{array}{c} \eta_q \\ \eta_s\end{array} \right)
= \left(\begin{array}{cc}
\cos\phi & -\sin\phi\\ \sin\phi & \cos\phi\\ \end{array} \right)
\left(\begin{array}{c} \eta_q \\ \eta_s\end{array} \right),\label{eq:e-ep2}
\eeq
where $\phi$ is the mixing angle.
The relation between the decay constants $(f_\eta^q, f_\eta^s,f_{\etar}^q,f_{\etar}^s)$ and
$(f_q,f_s)$ can be found in Refs.~\cite{zjx2008,fks98}.
The chiral enhancements $m_0^q$ and $m_0^s$ have been defined
in Ref.~\cite{nlo05} by assuming the exact isospin symmetry $m_q=m_u=m_d$.
The three input parameters $f_q, f_s,$ and $\phi$ in the FKS mixing scheme have been
extracted from the data of the relevant exclusive processes \cite{fks98}
\beq
f_q=(1.07\pm 0.02)f_{\pi},\quad f_s=(1.34\pm 0.06)f_{\pi},\quad \phi=39.3^\circ
\pm 1.0^\circ.
\eeq
With $f_\pi=0.13$ GeV, the chiral enhancements $m_0^q$ and $m_0^s$ consequently
take the values of $m_0^q=1.07$ GeV and $m_0^s=1.92$ GeV \cite{nlo05}.

In Ref.~\cite{eepg}, the authors extended the conventional FKS mixing scheme to include
the possible pseudoscalar glueball $G$, i.e., a small gluonic component
in both $\eta$ and $\etar$ mesons \cite{en2007}. In their $\eta$-$\etar$-$G$ mixing
scheme, the physical
states $(\eta, \eta', G)$ are related to $(\eta_8,\eta_1,g)$ and $(\eta_q,\eta_s,g)$
through the rotation $U(\phi,\phi_G)=U_3(\theta) U_1(\phi_G) U_3(\theta_i)$\cite{eepg},
\beq
\label{eq:qs2}
 \left( \begin{array}{c}    |\eta\rangle \\ |\eta'\rangle\\|G\rangle
   \end{array} \right)   = U_3(\theta)U_1(\phi_G)\left( \begin{array}{c}
    |\eta_8\rangle \\ |\eta_1\rangle\\|g\rangle   \end{array} \right)
    =U_3(\theta) U_1(\phi_G) U_3(\theta_i)\left( \begin{array}{c}
    |\eta_q\rangle \\ |\eta_s\rangle\\|g\rangle   \end{array} \right) \;,
\eeq
with the rotation matrices \cite{eepg}
\beq
U_3(\theta)&=&\left( \begin{array}{ccc}
\cos\theta & -\sin\theta & 0\\ \sin\theta & \cos\theta & 0\\ 0 &0&1   \end{array} \right)\;,
\quad
U_1(\phi_G )=\left( \begin{array}{ccc}
1&0&0\\
0&\cos\phi_G & \sin\phi_G \\
0&-\sin\phi_G & \cos\phi_G\\  \end{array} \right)\;,
\non
U(\phi,\phi_{G})&=&\left( \begin{array}{ccc}
\cos\phi+\sin\theta\sin\theta_i\Delta_G & -\sin\phi+\sin\theta\cos\theta_i\Delta_G & -\sin\theta\sin\phi_G\\
\sin\phi-\cos\theta\sin\theta_i\Delta_G & \cos\phi-\cos\theta\cos\theta_i\Delta_G & \cos\theta\sin\phi_G\\
  -\sin\theta_i\sin\phi_G &-\cos\theta_i\sin\phi_G&\cos\phi_G
   \end{array} \right)\;,\label{eq:mut}
\eeq
where $\theta_i=54.7^\circ$ is the ideal mixing angle with $\cos\theta_i=1/\sqrt{3}$ and
$\sin\theta_i=\sqrt{2/3}$, the angle $\phi=\theta+\theta_i$ and the
abbreviation $\Delta G=1-\cos\phi_G$. One can see that the matrix $U(\phi,\phi_{G})$
will approach the FKS mixing matrix \cite{fks98} in the limit of $\phi_G\to 0$, which
means that the angle $\phi$ in Eq.~(\ref{eq:mut}) plays the same role as the mixing angle
in the FKS mixing scheme\cite{eepg}.

The chiral masses $m_0^q$ and $m_0^s$ in the $\eta$-$\etar$-$G$ mixing scheme can be
written as \cite{eepg}
\beq
m^{q}_{0}&=&\frac{m^2_{qq}}{2m_q}=\frac{1}{2m_q}(U_{11} -\frac{\sqrt{2}f_s}{f_q}U_{12}),
\label{eq:m0q1}\\
m^{s}_{0}&=&\frac{m^2_{ss}}{2m_s}=\frac{1}{2m_s}(U_{22}-\frac{f_q}{\sqrt{2}f_s}U_{21}),\label{eq:m0s1}
\eeq
with the rotation matrix elements $U_{ij}$ having the form \cite{eepg}
\beq
U_{11}&=&(\cos\phi+\sin\theta \sin\theta_i \Delta_G)^2 m_\eta^2 +
(\sin\phi-\cos\theta \sin\theta_i\Delta_G)^2 m_{\eta'}^2 + (\sin\theta_i \sin\phi_G)^2 m_G^2\;,\non
U_{12}&=&U_{21}= (\cos\phi+\sin\theta \sin\theta_i \Delta_G) \cdot
(-\sin\phi+\sin\theta \cos\theta_i \Delta_G) m_\eta^2 \non && +
(\sin\phi-\cos\theta \sin\theta_i \Delta_G)\cdot
(\cos\phi-\cos\theta \cos\theta_i\Delta_G) m_{\eta'}^2
 + \sin\theta_i \sin\phi_G \cdot \cos\theta_i \sin\phi_G m_G^2\;, \non
U_{22}&=&(-\sin\phi+\sin\theta \cos\theta_i \Delta_G)^2 m_\eta^2 +
(\cos\phi-\cos\theta \cos\theta_i \Delta_G)^2 m_{\eta'}^2\non && + (\cos \theta_i \sin \phi_G )^2 m_G^2\;.
\label{eq:uij}
\eeq
The decay constants associated with the physical $(\eta,\eta',G)$
states are related to those associated with
$(\eta_q,\eta_s, g)$ states via the same mixing matrix\cite{eepg}
\beq
\left( \begin{array}{cc}f_\eta^q & f_\eta^s \\ f_{\eta'}^q & f_{\eta'}^s \\
f_G^q &f_G^s \end{array} \right)=
U(\phi,\phi_G) \left( \begin{array}{cc} f_q & f_q^s \\ f_s^q & f_s \\ f_g^q & f_g^s
\end{array} \right)\;. \label{eq:fpi}
\eeq
The mixing angle $\phi_G$ describes the mixing between the flavor singlet
$\eta_1$ and unmixed glueball $g$ and can vary in a range, depending on the
parametrization of the mixing matrix, experimental inputs, and the fitting procedure.
Since current data and known theoretical estimations \cite{li2008}
suggest a rather small gluonic component in $\etap$, the angle $\phi_G$
should be small as well. Following Ref.~\cite{eepg}, we also take $\phi_G=12^\circ$.

For the mass of the pseudoscalar glueball, the theoretical prediction for $m_G$
depends on the choice of input parameters, such as $(m_\eta,m_{\etar},f_s,f_q, etc)$.
The theoretical estimations in Refs.~\cite{eepg,eepgec} 
prefer a value of $m_G \approx 1.3-1.5 $ GeV. If we take
$f_q=f_\pi$ and $f_s=1.3 f_\pi$  in the numerical estimations
, we find that $m_G=1.376$ GeV ( see Table I of Ref.~\cite{eepgec} ) from the
approximate correlation relation between $m_G$ and the other input parameters
as given in Eq.(35) of Ref.~\cite{eepgec}.
For other input parameters we also follow the choice of Refs.~\cite{eepg,eepgec},
and finally we take
\beq
f_q=f_\pi, \quad f_s=1.3 f_\pi, \quad \phi=43.7^\circ
\quad \phi_G=12^\circ, \quad m_G=1.376 GeV,
\label{eq:ms2-input}
\eeq
in the numerical calculations when the $\eta$-$\etar$-$G$ scheme is adopted.

In the third $\eta$-$\etar$-$G$-$\eta_c$ mixing scheme\cite{eepgec},
finally, the flavor states
are transformed into the physical states through the mixing
matrix $U(\theta,\phi_G,\phi_Q)$:
\beq
\left( \begin{array}{c}
    |\eta\rangle \\ |\eta'\rangle\\|G\rangle\\|\eta_c\rangle
   \end{array} \right)
   = U(\theta,\phi_G,\phi_Q)
\left( \begin{array}{c}
    |\eta_q\rangle \\ |\eta_s\rangle\\|g\rangle\\|\eta_Q\rangle
   \end{array} \right),
\eeq
with the $4\times 4$ mixing matrix \cite{eepgec}
\beq
U(\theta,\phi_G,\phi_Q)=\left(
\begin{array}{cccc}
c\theta c\theta_i-s\theta c\phi_G s\theta_i & -c\theta
s\theta_i-s\theta c\phi_G c\theta_i
& -s\theta s\phi_G c\phi_Q& -s\theta s\phi_G s\phi_Q\\
s\theta c\theta_i+c\theta c\phi_G s\theta_i & -s\theta
s\theta_i+c\theta c\phi_G c\theta_i & c\theta s\phi_G c\phi_Q& c\theta s\phi_G s\phi_Q\\
-s\phi_G s\theta_i &-s\phi_G c\theta_i & c\phi_G
c\phi_Q & c\phi_G s\phi_Q \\
0 &0 & -s\phi_Q & c\phi_Q
 \end{array} \right),\label{eq:mut4}
\eeq
where $\theta_i=54.7^\circ$ is the ideal mixing angle, $\theta=\phi-\theta_i$
(here $\phi$ is the previous mixing angle in the FKS mixing scheme), and
$c\theta$ ($s\theta$) is the abbreviation for $\cos\theta$
($\sin\theta$) and similarly for others.
With the definition of $\phi=\theta+\theta_i$, the rotation matrix
$U(\theta,\phi_G,\phi_Q)$ in Eq.~(\ref{eq:mut4}) approaches the
mixing matrix $U(\phi,\phi_G)$ in Eq.~(\ref{eq:mut}) in the $\phi_Q\to 0$ limit,
or the FKS mixing matrix $U(\phi)$ \cite{fks98} in the limits
of $\phi_Q\to 0$ and $\phi_G\to 0$.

The decay constants associated with the $\eta$, $\eta'$, $G$, and $\eta_c$ physical
states are related to those associated with the $\eta_q$, $\eta_s$,
$g$, $\eta_Q$ states through the mixing matrix  $U(\theta,\phi_G,\phi_Q)$:
\beq
\left(\begin{array}{ccc}
f_\eta^q & f_\eta^s & f_\eta^c \\
f_{\eta'}^q & f_{\eta'}^s & f_{\eta'}^c \\
f_G^q &f_G^s &f_G^c \\
f_{\eta_c}^q & f_{\eta_c}^s & f_{\eta_c}^c
\end{array} \right)=
U(\theta,\phi_G,\phi_Q) \left(
\begin{array}{ccc}
f_q & f_q^s & f_q^c\\
f_s^q & f_s & f_s^c \\
f_g^q & f_g^s & f_g^c \\
f_c^q & f_c^s & f_c
\end{array} \right).\label{fpi2}
\eeq
Furthermore, we find that the chiral masses $m_0^q$ and $m_0^s$ 
in this mixing scheme are identical to those as
given in Eqs.~(\ref{eq:m0q1}) and (\ref{eq:m0s1}).
In the $\eta$-$\etar$-$G$-$\eta_c$ mixing scheme, we also take $\phi_Q=11^\circ $ 
as per Ref.~\cite{eepgec},  while we choose the other input parameters to be 
the same as those given in Eq.~(\ref{eq:ms2-input}), i.e.,
\beq
f_q=f_\pi, \quad f_s=1.3 f_\pi, \quad \theta=-11^\circ,
\quad \phi_G=12^\circ, \quad \phi_Q=11^\circ, \quad m_G=1.376 GeV,
\label{eq:ms3-input}
\eeq

\subsection{ Wave functions}\label{sec:2-1b}

The $B$ meson is treated as a very good heavy-light system.
Its wave function can be written as the form of
\beq
\Phi_B= \frac{i}{\sqrt{2N_c}} (\psl_B +m_B) \gamma_5 \phi_B ({\bf k_1}).
\label{bmeson}
\eeq
Here we have adopted the B-meson distribution amplitude  $\phi_B(x,b)$ widely used
in the studies of B-meson hadronic decays based on the pQCD
factorization approach since 2001 \cite{yy01,kls01,lu2001,plb504,prd65,li2003}
\beq
\phi_B(x,b)&=& N_B x^2(1-x)^2 \exp \left
 [ -\frac{M_B^2\ x^2}{2 \omega_{b}^2} -\frac{1}{2} (\omega_{b} b)^2\right],
 \label{phib}
\eeq
where the $b$-dependence was included through the second term in the exponential
function, the shape parameter $\omega_b =0.40\pm 0.04$ has been fixed \cite{li2003}
from the fit to the $B \to \pi$ form factors derived from lattice QCD \cite{plb486}
and from the light-cone sum rule \cite{ball98} 
and finally the normalization factor $N_B$ depends on the values of
$\omega_b$ and $f_B$ and defined through the normalization relation
\beq
\int_0^1dx \; \phi_B(x,b=0)=\frac{f_B}{2\sqrt{6}}.
\eeq

The wave functions of the final-state mesons $M=(K,\eta_q,\eta_s)$ are defined as:
\beq
\Phi_{M_i}(P_i,x_i)\equiv \frac{i}{\sqrt{2N_C}}\gamma_5 \left [
\psl_i \phi_{M_i}^{A}(x_i)+m_{0i} \phi_{M_i}^{P}(x_i)+ m_{0i} (\nsl\vsl - 1)\phi_{M_i}^{T}(x_i)\right ],
\eeq
where $m_{0i}$ is the corresponding meson chiral mass, and $P_i$ and $x_i$ are the momentum
and the momentum fraction of $M_i$, respectively.
The explicit expressions of the distribution amplitudes $\phi_{M_i}^{A,P,T}(x_i)$ for
$M=(K, \eta_q,\eta_s)$ are given in Appendix A.

In the third $\eta$-$\etar$-$G$-$\eta_c$ mixing scheme, the $\eta_c$ part of the $\etap$ meson will contribute to
the $B \to K \etap$ decays through the decay chain $B \to K \eta_c \to K \etap$.
The wave function of the $\eta_c$ can be written as\cite{plb612}:
\beq
\Phi_{\eta_c}(P_2,x_2) \equiv \frac{i}{\sqrt{2N_C}} \gamma_5[
\psl_2 \phi_{\eta_c}^{\nu}(x_2)+m_{\eta_c} \phi_{\eta_c}^{s}(x_2)].
\eeq
The distribution amplitudes $\phi_{\eta_c}^{\nu, s}$ are of the form \cite{plb612}:
\begin{eqnarray}
\phi_{\eta_c}^{\nu}(x) &=& 9.58\frac{f_{\eta_c}}{2\sqrt{2N_c}}x(1-x)\Big[\frac{x(1-x)}{1-2.8x(1-x)}\Big]^{0.7},\non
\phi_{\eta_c}^s(x) &=& 1.97\frac{f_{\eta_c}}{2\sqrt{2N_c}}\Big[\frac{x(1-x)}{1-2.8x(1-x)}\Big]^{0.7}.
\end{eqnarray}

\section{$B \to K \etap$ and $B \to K \eta_c$ decays at leading order}\label{sec:lo-1}

\begin{figure}[tb]
\vspace{-5cm}
\centerline{\epsfxsize=18cm \epsffile{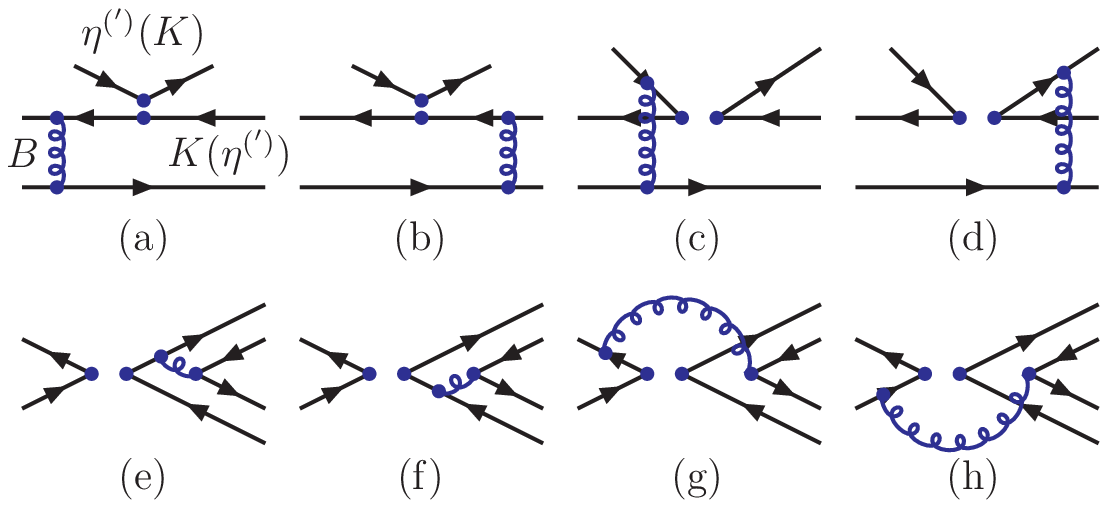}}
\vspace{-15cm}
\caption{ Feynman diagrams that may contribute to the
$B\to K \etap$ decays in the pQCD approach at leading order.}
\label{fig:fig1}
\end{figure}

In this section we will present the total decay amplitudes for $B \to K \etap$ and $B \to K \eta_c$ decays
in the pQCD approach at leading order.

\subsection{$B \to K \etap$ decays at leading order }

The $B \to K \etap$ decays have been studied 
by using the ordinary $\eta$-$\etar$ mixing scheme and by 
employing the pQCD factorization approach at
the LO and partial NLO level in Ref.~\cite{zjx2008}.
We recalculated and confirmed the relevant analytical formulas as given in Ref.~\cite{zjx2008}.
For the sake of the reader, we here present directly the decay amplitudes obtained by
evaluating the Feynman diagrams Figs.1(a)-1(h).

For the factorizable emission diagrams Figs.1(a) and 1(b), the decay amplitudes 
for the cases of a $B\to K$ transition are of the form
\beq
F^{V-A}_{eK}&=&-F^{V+A}_{eK} = -8\pi
C_FM_{B}^4\int^1_0dx_1dx_3\int^\infty_0b_1db_1b_3db_3
\phi_{B}(x_1,b_1)
  \nonumber\\
  &&\times\Big\{ \Big[(1+x_3)\phi_K^A(x_3)
  +r_3(1-2x_3)(\phi_K^P(x_3)+\phi_K^T(x_3))
  \Big]\nonumber\\
  &&\times E_e(t_a)h_e(x_1,x_3,b_1,b_3)
+2r_3\phi_K^P(x_3)E_e(t_a^\prime)h_e(x_3,x_1,b_3,b_1)
 \Big\},\label{ppefll}\\
F^{SP}_{eK}&=& -16\pi r_2
  C_FM_{B}^4\int^1_0dx_1dx_3\int^\infty_0b_1db_1b_3db_3
\phi_{B}(x_1,b_1)
  \nonumber\\
  &&\times\Big\{\Big[\phi_K^A(x_3)+r_3(2+x_3)\phi_K^P(x_3)-r_3x_3\phi_K^T(x_3)\Big]\nonumber\\
  &&\times E_e(t_a)
  h_e(x_1,x_3,b_1,b_3)
  +2  r_3\phi_K^P(x_3)E_e(t_a^\prime)h_e(x_3,x_1,b_3,b_1)\Big\},\label{ppefsp}
  \end{eqnarray}
where $C_F=4/3$, $r_2=r_\eta=m^{q,s}_{0}/M_B$, and $r_3=r_K=m^{K}_{0}/M_B$.
The hard functions, $E_e(t)$ and $h_e$, and the hard scales $t$ are given in Appendix B.

For the non-factorizable emission diagrams Figs.1(c) and 1(d), whose contributions are
\beq
M_{eK}^{V-A}&=&-\frac{32\pi C_FM_{B}^4}{\sqrt{6}}\int^1_0dx_1dx_2dx_3\int^\infty_0b_1db_1b_2db_2
\phi_{B}(x_1,b_1)\phi_\eta^A(x_2) \non
&&\times
\Big\{\Big[(1-x_2)\phi_K^A(x_3)-r_3x_3(\phi_K^P(x_3)-\phi_K^T(x_3))\Big]
E_e^\prime(t_b)h_n(x_1,\bar{x}_2,x_3,b_1,b_2)\nonumber\\
&&+\Big[-(x_2+x_3)\phi_K^A(x_3)+r_3x_3(\phi_K^P(x_3)+\phi_K^T(x_3))\Big]\non
 &&\times
 E_e^\prime(t_b^\prime)h_n(x_1,x_2,x_3,b_1,b_2)\Big\},\label{ppenll}
 \eeq
\beq
M_{eK}^{V+A}&=&-\frac{32\pi C_FM_{B}^4r_2}{\sqrt{6}}
      \int^1_0dx_1dx_2dx_3\int^\infty_0b_1db_1b_2db_2\phi_{B}(x_1,b_1)\non
     &&\times \Big\{\Big[(1-x_2)\phi_K^A(x_3)
     \left(\phi_\eta^P(x_2)+\phi_\eta^T(x_2)\right) +r_3x_3\left(\phi_\eta^P(x_2)-\phi_\eta^T(x_2)\right)
      \left(\phi_K^P(x_3)+\phi_K^T(x_3)\right)\non
&&  +(1-x_2)r_3\left(\phi_2^P(x_2)+\phi_2^T(x_2)\right)\left(\phi_3^P(x_3)
      -\phi_3^T(x_3)\right)\Big] E_e^\prime(t_b) h_n(x_1,\bar{x}_2,x_3,b_1,b_2)\non
&&-\Big[x_2\phi_K^A(x_3)(\phi_\eta^P(x_2)-\phi_\eta^T(x_2))
       +r_3x_2\left ( \phi_\eta^P(x_2)-\phi_\eta^T(x_2) \right )\left (\phi_K^P(x_3)-\phi_K^T(x_3) \right )\non
&& +r_3x_3\left (\phi_\eta^P(x_2) +\phi_\eta^T(x_2) \right )\left (\phi_K^P(x_3)+\phi_K^T(x_3) \right )\Big]
\cdot E_e^\prime(t_b^\prime)h_n(x_1,x_2,x_3,b_1,b_2)\Big\},\label{ppenlr}
\\
M^{SP}_{eK} &=&-\frac{32\pi C_F M_{B}^4}{\sqrt{6}}\int^1_0dx_1dx_2dx_3\int^\infty_0b_1db_1b_2db_2
\phi_{B}(x_1,b_1)\phi_\eta^A(x_2) \non
&&\times \Big\{ \Big[(x_2-x_3-1)\phi_K^A(x_3)+r_3x_3(\phi_K^P(x_3)+\phi_K^T(x_3))\Big]
E_e^\prime(t_b)h_n(x_1,\bar{x}_2,x_3,b_1,b_2)\non
&&+\Big[x_2\phi_K^A(x_3)+r_3x_3(\phi_K^T(x_3)-\phi_K^P(x_3))\Big]
E^\prime_e(t_b^\prime)h_n(x_1,x_2,x_3,b_1,b_2)\Big\},\label{ppensp}
\eeq
where $\bar{x}_i=1-x_i$, while $\phi_{\eta}$ denotes $\phi_{\eta_q}$ or $\phi_{\eta_s}$.

For the factorizable annihilation diagrams Figs.1(e) and 1(f), 
the decay amplitudes are of the form
\beq
 F_{aK}^{V-A}&=& F_{aK}^{V+A}= -8\pi
C_FM_{B}^4\int^1_0dx_2dx_3\int^\infty_0b_2db_2b_3db_3\times \Big\{\Big[(x_3-1)\phi_\eta^A(x_2)\phi_K^A(x_3)\non
&& -4r_2r_3\phi_\eta^P(x_2)\phi_K^P(x_3)
+2r_2r_3x_3\phi_\eta^P(x_2)(\phi_K^P(x_3)-\phi_K^T(x_3))\Big] E_a(t_c)h_a(x_2,\bar{x}_3,b_2,b_3)\non
&&+\Big[x_2\phi_\eta^A(x_2)
\phi_K^A(x_3)+2r_2r_3(\phi_\eta^P(x_2)-\phi_\eta^T(x_2))\phi_K^P(x_3)
\non
&&+2r_2r_3x_2(\phi_\eta^P(x_2)+\phi_\eta^T(x_2))\phi_K^P(x_3)\Big]
 E_a(t_c^\prime)h_a(\bar{x}_3,x_2,b_3,b_2)\Big\},\label{ppafll}\\
F_{aK}^{SP}&=&-16\pi
  C_FM_{B}^4\int^1_0dx_2dx_3\int^\infty_0b_2db_2b_3db_3\non
  &&\hspace{-1cm}\times \Big\{\Big[2r_2\phi_\eta^P(x_2)\phi_K^A(x_3)+(1-x_3)r_3\phi_\eta^A(x_2)(\phi_K^P(x_3)
  +\phi_K^T(x_3))\Big] E_a(t_c)h_a(x_2,\bar{x}_3,b_2,b_3)\non
  &&\hspace{-1cm}+\Big[2r_3\phi_\eta^A(x_2)\phi_K^P(x_3)+r_2x_2(\phi_\eta^P(x_2)-\phi_\eta^T(x_2))\phi_K^A(x_3)  \Big]
  E_a(t_c^\prime)h_a(\bar{x}_3,x_2,b_3,b_2)\Big\}.\label{ppafsp}
\eeq

The contributions from the non-factorizable annihilation diagrams,
Figs.1(g) and 1(h), are
\beq
M_{aK}^{V-A}&=&-\frac{32\pi C_FM_{B}^4}{\sqrt{6}}\int^1_0dx_1dx_2dx_3\int^\infty_0b_1db_1 b_2db_2\phi_{B_s}(x_1,b_1)\nonumber\\
 &&\hspace{-1cm}\times \Big\{\Big[-x_2\phi_\eta^A(x_2)\phi_K^A(x_3)-4r_2r_3
 \phi_\eta^P(x_2)\phi_K^P(x_3)\non
 &&\hspace{-1cm}+r_2r_3(1-x_2)(\phi_\eta^P(x_2)+\phi_\eta^T(x_2))(\phi_K^P(x_3)-\phi_K^T(x_3))
 \non
 &&\hspace{-1cm}+r_2r_3x_3(\phi_\eta^P(x_2)-\phi_\eta^T(x_2))(\phi_K^P(x_3)+\phi_K^T(x_3))\Big]
 E_a^\prime(t_d)h_{na}(x_1,x_2,x_3,b_1,b_2)\non
 &&\hspace{-1cm}+\Big[\bar{x}_3\phi_\eta^A(x_2)\phi_K^A(x_3)
 +\bar{x}_3 r_2r_3(\phi_\eta^P(x_2)+\phi_\eta^T(x_2))(\phi_K^P(x_3)-\phi_K^T(x_3))
 \non
 && \hspace{-1cm}+x_2r_2r_3(\phi_\eta^P(x_2)-\phi_\eta^T(x_2))(\phi_K^P(x_3)+\phi_K^T(x_3))\Big]
 E_a^\prime(t_d^\prime)h_{na}^\prime(x_1,x_2,x_3,b_1,b_2)\Big\},\label{ppanll}
 \eeq
\beq
M_{aK}^{V+A}&=&-\frac{32\pi C_FM_{B_s}^4}{\sqrt{6}}\int^1_0dx_1dx_2dx_3\int^\infty b_1db_1b_2db_2\phi_{B}(x_1,b_1)\non
 &&\times \Big\{\Big[r_2(2-x_2)(\phi_\eta^P(x_2)+\phi_\eta^T(x_2))
 \phi_K^A(x_3)\non
 &&-r_3(1+x_3)\phi_\eta^A(x_2)(\phi_K^P(x_3)-\phi_K^T(x_3))\Big]
  E_a^\prime(t_d)h_{na}(x_1,x_2,x_3,b_1,b_2) \non
 &&+\Big[r_2x_2\left(\phi_\eta^P(x_2)+\phi_\eta^T(x_2)\right)\phi_K^A(x_3)
  -r_3 \bar{x}_3 \phi_\eta^A(x_2) (\phi_K^P(x_3)-\phi_K^T(x_3))\Big] \non
 &&\times E_a^\prime(t_d^\prime)h_{na}^\prime (x_1,x_2,x_3,b_1,b_2)
 \Big\},\label{ppanlr}
\eeq
\beq
M_{aK}^{SP}&=&-\frac{32\pi C_F M_{B}^4}{\sqrt{6}}\int^1_0dx_1dx_2dx_3\int^\infty_0b_1db_1b_2db_2
 \phi_{B}(x_1,b_1)\non
 &&\times \Big\{\Big[(x_3-1)
 \phi_\eta^A(x_2)\phi_K^A(x_3)-4r_2r_3\phi_\eta^P(x_2)\phi_K^P(x_3)
 \non
 &&+r_2r_3x_3(\phi_\eta^P(x_2)+\phi_\eta^T(x_2)) (\phi^P_K(x_3)-\phi_K^T(x_3))\non
 &&+r_2r_3(1-x_2)(\phi_\eta^P(x_2)-\phi_\eta^T(x_2))(\phi^P_K(x_3)+\phi_K^T(x_3))\Big]
 E_a^\prime(t_d)h_{na}(x_1,x_2,x_3,b_1,b_2) \non
 &&+\Big[x_2\phi_\eta^A(x_2)\phi_K^A(x_3)+x_2r_2r_3(\phi_\eta^P(x_2)+\phi_\eta^T(x_2))
 (\phi_K^P(x_3)-\phi_K^T(x_3))\non
 &&+r_2r_3(1-x_3)(\phi_\eta^P(x_2)-\phi_\eta^T(x_2))(\phi_K^P(x_3)+\phi_K^T(x_3))\Big]\non
 &&\times
 E_a^\prime(t_d^\prime)h_{na}^\prime(x_1,x_2,x_3,b_1,b_2)\Big\},
 \label{ppansp}
 \end{eqnarray}
The evolution functions $E_i(t)$ and hard functions $h_i$ appearing in
Eqs.~(\ref{ppefll})-(\ref{ppansp}) are given explicitly in Appendix B.

If we exchange the position of $K$ and $\etap$ in Fig.~1, we will find the corresponding
decay amplitudes for the case of $B \to \etap$ transitions.
Since the $K$ and $\etap$ are all
pseudoscalar mesons and have the similar wave functions, the decay
amplitudes for new diagrams $-$ say $F_{e\eta}^{V-A}$,
$F_{e\eta}^{V+A}$, $F_{e\eta}^{SP}$, $M_{e\eta}^{V-A}$,
$M_{e\eta}^{V+A}$, $M_{e\eta}^{SP}$, $F_{a\eta}^{V-A}$,
$F_{a\eta}^{V+A}$, $F_{a\eta}^{SP}$, $M_{a\eta}^{V-A}$,
$M_{a\eta}^{V+A}$, and $M_{a\eta}^{SP}$ $-$ can be obtained from those
as given in Eqs.~(\ref{ppefll})-(\ref{ppansp}) by the following
replacements:
\beq
\pka \leftrightarrow \phi^{A}_{\eta^{(\prime)}},
\quad \pkp \leftrightarrow \phi^{P}_{\eta^{(\prime)}} , \quad \pkt
\leftrightarrow \phi^{T}_{\eta^{(\prime)}}, \quad r_\eta
\leftrightarrow r_K. \label{repalce1}
\eeq

\begin{figure}[tb]
\vspace{-6cm}
\begin{center}
\leftline{\epsfxsize=18cm\epsffile{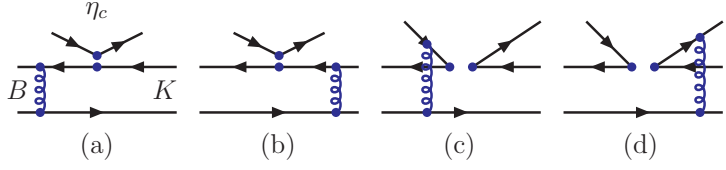}}
\end{center}
\vspace{-19cm}
\caption{Feynman diagrams which may contribute to the $B\to K \eta_c$ decays in 
the pQCD approach at leading order.}
\label{fig:fig2}
\end{figure}

\subsection{$B \to K \eta_c$ at leading order}

For $B \to K \eta_{c}$ decays at leading order in the pQCD approach, 
the Feynman diagrams which may contribute
are shown in Fig.~\ref{fig:fig2}. The $B \to K \eta_{c}$ decays have been studied
in Ref.~\cite{liux10} in the pQCD approach, at the full leading order and with the
inclusion of the partial NLO vertex corrections. We here recalculated theses decays
and confirmed the results of Ref.~\cite{liux10}. The relevant decay amplitudes are the following
\beq
F_{\eta_c K}&=&-8\pi C_FM_{B}^4 \int^1_0dx_1dx_3\int^\infty_0b_1db_1b_3db_3
\phi_{B}(x_1,b_1)  \non
  && \times \Big\{ \Big[[(1-r_{\eta_c}^2)(1+x_3)-x_3r_{\eta_c}^2]\phi_K^A(x_3)+r_3(1-2x_3)[\phi_K^P(x_3)+\phi_K^T(x_3)] \non
  && +r_3 r_{\eta_c}^2[(1+2x_3)\phi_K^P(x_3) - (1-2x_3)\phi_K^T(x_3)]
  \Big]E_e(t_e)h_e^\prime(x_1,x_3,b_1,b_3) \non
 && +2r_3(1-r_{\eta_c}^2)\phi_K^P(x_3)E_e(t_e^\prime)h_e^\prime(x_3,x_1,b_3,b_1) \Big\},
 \label{ppefllec}\\
M_{\eta_c K} &=& \frac{32}{\sqrt{6}}\pi C_FM_{B}^4\int^1_0dx_1dx_2dx_3\int^\infty_0b_1db_1b_2db_2
\phi_{B}(x_1,b_1) \phi_{\eta_c}^{\nu}(x_2) \non
&& \times  \Big[x_3(1-2r_{\eta_c}^2)\phi_K^A(x_3) - 2x_3 r_3(1-r_{\eta_c}^2)\phi_K^T(x_3)\Big]\non
&& \ \ \ \cdot E_e^\prime(t_f^\prime)h_n^\prime(x_1,x_2,x_3,b_1,b_2),
\label{ppenllec}
\eeq
where $r_3=m^K_0/m_B$, $r_{\eta_c}=m_{\eta_c}/m_B$, and $r_c=m_c/m_B$.
$\phi_{\eta_c}^{\nu}$ is the leading twist-2 part of the distribution amplitude for the
pseudoscalar meson $\eta_c$. The evolution function
$E^{(\prime)}(t)$, hard function $h_i$ and the scale $t_e$,
$t_e^\prime$ are given in Appendix B.

In the leading-order pQCD approach, the total decay amplitude for the 
$B \to \eta_c K$ decay can then be written as:
\beq
{\cal M}(B \to \eta_c K) &=& \frac{G_F}{\sqrt{2}}F_{\eta_c K} f_{\eta_c}\Big[V_{cb}^*V_{cs} a_2 -
V_{tb}^*V_{ts}(a_3-a_5-a_7+a_9)\Big] \nonumber\\
  &&+ \frac{G_F}{\sqrt{2}}M_{\eta_c K}\Big[V_{cb}^*V_{cs}C_2 - V_{tb}^*V_{ts}
  (C_4+C_6+C_8+C_{10})\Big],
\eeq
where $a_i$ is the combination of the Wilson coefficients
$C_i$:
\beq
a_{1,2}&=&C_{2,1}+\frac{C_{1,2}}{3}; \quad
a_i=C_i+\frac{C_{i+1}}{3}, \quad for \quad i=3,5,7,9;\non
a_i&=&C_i+\frac{C_{i-1}}{3}, \quad for \quad i=4,6,8,10.
\label{eq:ai}
\eeq

\subsection{ Total decays amplitudes of $B \to K \etap$ decays}

For $B^0 \to K^0 \eta$ and $B^+\to K^+\eta$ decays, by combining the contributions from
all possible Feynman diagrams ( Figs.~\ref{fig:fig1} and
\ref{fig:fig2}), one finds the general expressions for the
total decay amplitudes (here the Wilson coefficients and Cabibbo-Kobayashi-Maskawa (CKM)
factors are all included),
\beq
{\cal M}(K^0 \eta) &=& \frac{G_F}{\sqrt{2}}\Big\{\lambda_u\Big[a_2f_{\eta}^qF_{eK}^{V-A}+C_2M_{eK}^{V-A,q}F_1(\phi)\Big]
- \lambda_t\left (2a_3-2a_5-\frac{1}{2}a_7+\frac{1}{2}a_9\right )f_{\eta}^qF_{eK}^{V-A}\non
&&
- \lambda_t\Big[ \left (a_3+a_4-a_5+\frac{1}{2}a_7-\frac{1}{2}a_9-\frac{1}{2}a_{10}\right )f_{\eta}^sF_{eK}^{V-A}\non
&& +\left (a_6 -\frac{1}{2}a_8\right )
\left [f_{\eta}^sF_{eK}^{SP}+f_B F_{aK}^{SP}F_2(\phi)+ f_B F_{a\eta}^{SP}F_1(\phi)+f_KF_{e\eta}^{SP}F_1(\phi) \right]\non
&&+\left (C_3+C_4 -\frac{1}{2}C_{9}-\frac{1}{2}C_{10}\right )M_{eK}^{V-A,s}F_2(\phi)\non &&
+\left (2C_4+\frac{1}{2}C_{10}\right ) M_{eK}^{V-A,q}F_1(\phi)
+\left (C_5-\frac{1}{2}C_7\right )M_{eK}^{V+A,s}F_2(\phi)\non
&&
+\left (2C_6+\frac{1}{2}C_8\right )M_{eK}^{SP,q}F_1(\phi)+\left (C_6-\frac{1}{2}C_8\right )M_{eK}^{SP,s}F_2(\phi)\non
&&
+\left (a_4-\frac{1}{2}a_{10}\right )
\left [f_BF_{aK}^{V-A}F_2(\phi)+f_KF_{e\eta}^{V-A}F_1(\phi)+f_BF_{a\eta}^{V-A}F_1(\phi) \right]\non
&&
+\left (C_3-\frac{1}{2}C_9\right )
\left [M_{aK}^{V-A}F_2(\phi)+M_{e\eta}^{V-A}F_1(\phi)+M_{a\eta}^{V-A}F_1(\phi) \right]
\non
&&+\left (C_5-\frac{1}{2}C_7\right )
\left[ M_{aK}^{V+A}F_2(\phi)+M_{e\eta}^{V+A}F_1(\phi)+M_{a\eta}^{V+A}F_1(\phi) \right]\Big]\Big\}
\non && +{\cal M}(B \to \eta_cK)*F_c(\theta,\phi_G,\phi_Q),
\label{eq:k0eta}
\eeq
\beq
{\cal M}(K^+ \eta) &=&
 \frac{G_F}{\sqrt{2}}\Big\{ \lambda_u \Big[ a_2f_{\eta}^qF_{eK}^{V-A}+C_2M_{eK}^{V-A,q}F_1(\phi)+a_1 f_B\left[  F_{aK}^{V-A}F_2(\phi)+ F_{a\eta}^{V-A}F_1(\phi)\right]\non
&&+ a_1 f_K F_{e\eta}^{V-A}F_1(\phi) +C_1\left[
M_{aK}^{V-A}F_2(\phi)+M_{e\eta}^{V-A}F_1(\phi)+M_{a\eta}^{V-A}F_1(\phi)
\right]\Big]\non && - \lambda_t\Big[\left
(2a_3-2a_5-\frac{1}{2}a_7+\frac{1}{2}a_9\right
)f_{\eta}^qF_{eK}^{V-A}\non && +\left
(a_3+a_4-a_5+\frac{1}{2}a_7-\frac{1}{2}a_9-\frac{1}{2}a_{10}\right
)f_{\eta}^sF_{eK}^{V-A}\non && +\left (a_6 -\frac{1}{2}a_8\right
)f_{\eta}^sF_{eK}^{SP} + \left
(C_3+C_4-\frac{1}{2}C_{9}-\frac{1}{2}C_{10}\right
)M_{eK}^{V-A,s}F_2(\phi)\non && +\left (2C_4+\frac{1}{2}C_{10}\right
)M_{eK}^{V-A,q}F_1(\phi) +\left (C_5-\frac{1}{2}C_7\right
)M_{eK}^{V+A,s}F_2(\phi)\non && +\left (2C_6+\frac{1}{2}C_8\right
)M_{eK}^{SP,q}F_1(\phi) +\left (C_6-\frac{1}{2}C_8\right
)M_{eK}^{SP,s}F_2(\phi)\non && +(a_4+a_{10})\left
[f_BF_{aK}^{V-A}F_2(\phi)+f_KF_{e\eta}^{V-A}F_1(\phi)+f_BF_{a\eta}^{V-A}F_1(\phi)\right]\non
&&+(a_6+a_{8})\left [
f_BF_{aK}^{SP}F_2(\phi)+f_KF_{e\eta}^{SP}F_1(\phi)+f_BF_{a\eta}^{SP}F_1(\phi)
\right]\non && +(C_3+C_9)\left[
M_{aK}^{V-A}F_2(\phi)+M_{e\eta}^{V-A}F_1(\phi)+M_{a\eta}^{V-A}F_1(\phi)\right]\non
&&+(C_5+C_7)\left [
M_{aK}^{V+A}F_2(\phi)+M_{e\eta}^{V+A}F_1(\phi)+M_{a\eta}^{V+A}F_1(\phi)
\right]\Big]\Big\} \non && +{\cal M}(B \to
\eta_cK)*F_c(\theta,\phi_G,\phi_Q), \label{eq:kpeta}
\eeq
where $\lambda_u = V_{ub}^*V_{us}$, $\lambda_t =V_{tb}^*V_{ts}$, and the
Wilson coefficients $a_i$ are the same as those defined in Eq.~(\ref{eq:ai}).
The expressions for the mixing parameters $F_i^{(')}(\phi)$ depend on the choice of the different mixing
schemes:
\begin{enumerate}
\item[(i)]
In MS-1, i.e., the FKS $\eta$-$\etar$ mixing scheme, the mixing parameters
$F_{1,2}(\phi)$ and $F_{1,2}^\prime(\phi)$ are of the form
\beq
\sqrt{2}F_1(\phi)=F_2^\prime(\phi)=\cos(\phi), \quad
F_2(\phi)=-\sqrt{2}F_1^\prime(\phi)=-\sin(\phi).
\label{eq:fiphi1}
\eeq

\item[(ii)]
In MS-2, i.e., the $\eta$-$\etar$-$G$ mixing scheme, $F_{1,2}(\phi)$
and $F_{1,2}^\prime(\phi)$ are of the form
\beq
F_1(\phi)&=&
\frac{1}{\sqrt{2}}(\cos\phi+\sin\theta\sin\theta_i\Delta_G), \quad
F_2(\phi)= -\sin\phi+\sin\theta\cos\theta_i\Delta_G, \non
F_1^\prime(\phi)&=& \frac{1}{\sqrt{2}}(\sin\phi-\cos\theta\sin\theta_i\Delta_G), \quad
F_2^\prime(\phi)= \cos\phi-\cos\theta\cos\theta_i\Delta_G.
\label{eq:f1f2phig}
\eeq

\item[(iii)]
In MS-3, i.e., the third $\eta$-$\etar$-$G$-$\eta_c$ mixing scheme,
$F_{1,2}(\phi)$ and $F_{1,2}^\prime(\phi)$ are the same as those defined in Eq.~(\ref{eq:f1f2phig}).
For this case, the $\eta_c$ also contribute through mixing,
and the relevant mixing parameters are
\beq
F_c(\theta,\phi_G,\phi_Q)= -\sin\theta \sin\phi_G \sin\phi_Q,
\quad F_c^\prime(\theta,\phi_G,\phi_Q)= \cos\theta \sin\phi_G \sin\phi_Q.
\eeq

\end{enumerate}

Finally, the total decay amplitudes for $B^0 \to K^0 \etar$ and $B^+ \to K^+
\etar $ in the pQCD approach at leading order can be obtained easily from Eqs.~(\ref{eq:k0eta}) and
(\ref{eq:kpeta}) by the following replacements:
\beq
f_\eta^{q} \to f_{\eta^\prime}^q, \quad f_\eta^s \to f_{\eta^\prime}^s,\quad
F_1(\phi) \to F'_1(\phi), \quad F_2(\phi) \to F'_2(\phi), \quad F_c(\phi) \to F'_c(\phi).
\label{eq:f1f2phip}
\eeq

\section{NLO contributions in the pQCD approach}\label{sec:nlo}

In this section we will present the total decay amplitudes for $B \to K \etap$ decays
in the pQCD approach with the inclusion of all currently known NLO contributions.

\subsection{General analysis of the NLO contributions in the pQCD approach}

As is well known, the power-counting rule in the pQCD factorization approach
\cite{nlo05} is rather different
from that in the QCD factorization\cite{bn651,bbns99,npb675}.
When compared with the previous LO calculations in pQCD \cite{li2003},
the full pQCD predictions should include the following NLO contributions:

\begin{enumerate}
\item[(1)]
We should use the Wilson coefficients $C_i(\mw)$
at the NLO level in the naive dimensional regularization 
scheme \cite{buras96},
the NLO  renormalization group (RG) evolution
matrix $U(t,m,\alpha)$ as defined in Ref.~\cite{buras96}, and
the strong-coupling constant $\alpha_s(t)$ at the two-loop level.

\item[(2)]
Besides the LO hard kernel $H^{(0)}(\alpha_s)$, all the Feynman diagrams,
that contribute to $H^{(1)}(\alpha_s^2)$ in the pQCD
approach, should be considered. The typical Feynman diagrams which
contribute to $H^{(1)}(\alpha_s^2)$ at the NLO level in the pQCD approach are
shown in Figs. 3-5, and can be classified into six types.
\begin{itemize}
\item[(i)]
The vertex correction: the NLO contributions from the Feynman diagrams
as shown in Figs.3(a)-3(d), which were evaluated ten years ago\cite{bbns99,npb675,nlo05}.

\item[(ii)]
The quark loops: the NLO contributions from the quark loops as shown
in Figs. 3(e)-3(f), the relevant analytical formulas can be found in Ref.~\cite{nlo05}.

\item[(iii)]
The magnetic penguins: the NLO contributions from the operator $O_{8g}$,
as shown in Figs.3(g)-3(h). These Feynman diagrams
were evaluated several years ago \cite{o8g2003}.

\item[(iv)]
The NLO form factors(FF): i.e., the NLO contributions to the $B \to P$
transition form factors $F_0^{B\to P}(0)$ with $P=(K,\eta_q)$ in this paper.
The typical relevant Feynman diagrams are shown in Fig.~4, and were calculated
very recently in Ref.~\cite{prd85074004}.

\item[(v)]
The NLO contributions from the spectator diagrams as shown
in Figs.~5(a)-5(d), which are obtained by adding a new gluon line between any two
quark lines in Figs.~1(c) and 1(d), or by replacing the one-gluon lines with a three-gluon
vertex in Figs.~1(c) and 1(d). Such contributions are still unknown now.

\item[(vi)]
The NLO contributions from the annihilation diagrams, as shown by Figs.~5(e)-5(h),
which are obtained by adding a new gluon line between any two quark lines in
Figs.~1(e)-1(h).  Such contributions are also unknown now.

\end{itemize}

\end{enumerate}

The NLO contributions from the Feynman diagrams in Fig.~3 $-$
the vertex corrections, the quark-loops and chromo-magnetic penguins $-$
were evaluated several years ago \cite{bbns99,o8g2003,nlo05}, and taken into
account in our previous studies for the $B \to  K \etap$ decay in
Ref.~\cite{zjx2008}.

The Feynman diagrams as shown in Fig.~4 can provide the  NLO contributions to the
$B \to P$ transition form factors and have been calculated very recently
in Ref.~\cite{prd85074004}. The authors of Ref.\cite{prd85074004}
calculated the NLO corrections to the $B\to \pi$ transition form factors in
the leading twist in the $k_T$ factorization
theorem, and they found that the NLO part can provide a $ 20-30\%$ enhancement to the LO
results for the corresponding form factors. Since $\pi, K$ and $\etap$ are all
pseudoscalar mesons and have similar wave functions, it is straightforward 
to extend the
calculations in Ref.~\cite{prd85074004} to the cases for the $B \to K, \etap$ transition
form factors. In this paper, we will consider the effects of the NLO part of
the form factors. According to general expectations, the enhanced form factors
may lead to a larger branching ratio for $B \to K \etap$ decays.

The still-missing NLO parts in the pQCD approach are the $O(\alpha_s^2)$
contributions from nonfactorizable spectator diagrams and annihilation diagrams,
as illustrated by Figs.~5(a)-5(h).
The analytical calculations for these Feynman diagrams are still absent at present.
But it is generally believed that the NLO contributions from these Feynman diagrams
should be small
\begin{enumerate}
\item[(i)] The contributions from the nonfactorizable spectator 
diagrams in Figs.~1(c) and 1(d),
their contributions at leading order are strongly suppressed by the isospin symmetry
and color suppression with respect to the factorizable emission diagrams Figs.~1(a) and 1(b).
The NLO contributions from Figs.~5(a)-5(d) are higher-order effects on small LO quantities,
and therefore should be much smaller than the LO ones.

\item[(ii)]
The annihilation spectator diagrams at leading order, i.e., Figs.~1(e)-1(h),
they are power suppressed and generally much smaller with respect to the contributions from
the emission diagrams Figs.1(a) and 1(b). The contributions from Figs.~5(a)-5(d)
are also the higher-order corrections to the small quantities and
therefore should be much smaller than its LO parts.
\end{enumerate}

In the next section, we will explicitly evaluate the numerical values of the individual
decay amplitudes corresponding to different Feynman diagrams, and will compare the
size of every part of the total decay amplitude for the considered decays.
We will try to make a simple and clear comparison between the contributions from
different sets of Feynman diagrams or from the different sources numerically.

\begin{figure}[tb]
\vspace{-6cm}
\hspace{-3cm}\leftline{\epsfxsize=18 cm \epsffile{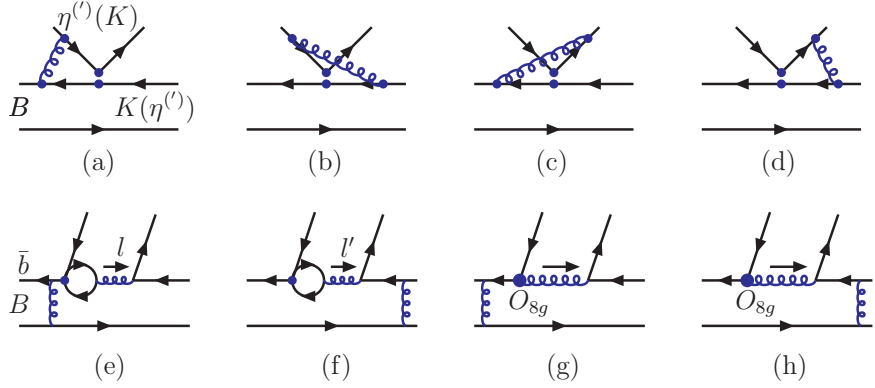}}
\vspace{-15.5cm}
\caption{The typical Feynman diagrams that provide NLO contributions to $B \to K \etap$ decays
in the pQCD approach: (a)-(d) are the vertex corrections, (e)-(f) are the 
quark-loops, and  (g)-(h) are the chromo-magnetic Penguins $O_{8g}$.}
\label{fig:fig3}
\end{figure}

\begin{figure}[h]
\vspace{-5.5cm}
\hspace{-3cm}\leftline{\epsfxsize=18 cm \epsffile{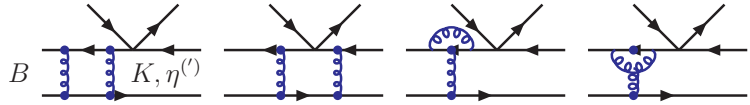}}
\vspace{-18.5cm}
\caption{The typical Feynman diagrams that may provide NLO contributions to
$B \to P$ form factors.}
\label{fig:fig4}
\end{figure}
\begin{figure}[h]
\vspace{-5.5cm}
\hspace{-3cm}\leftline{\epsfxsize=18 cm \epsffile{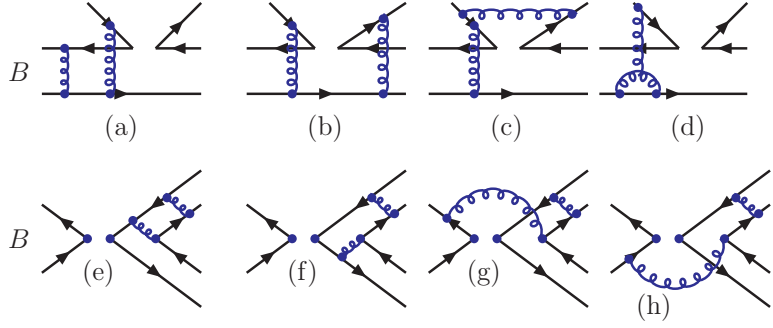}}
\vspace{-15.5cm}
\caption{The typical Feynman diagrams that may provide NLO contributions to
$B \to K \etap$ decays in the pQCD approach: (a)-(d) are the spectator
diagrams, (e)-(h) are the annihilation diagrams.}
\label{fig:fig5}
\end{figure}

\subsection{The NLO contributions to $B \to K \etap$ decays}\label{sec:vc1}

In Ref.~\cite{zjx2008}, by using the ordinary FKS $\eta-\etap$ mixing scheme,
we calculated the branching ratios and CP-violating
asymmetries of the four $B \to K \etap$ decays at leading order and partial
next-to-leading order, i.e., the NLO contributions from the Feynman diagrams in 
Fig~.3 were taken into account in Ref.~\cite{zjx2008}.
For details about the calculations of these NLO contributions and
the expressions of their relevant functions, we refer the reader to Ref.~\cite{zjx2008}.
For the sake of the reader, we here give a brief summary of  these ``old" NLO parts.

\begin{enumerate}

\item[(a)]{ {\bf Vertex corrections(VC):}}

The vertex corrections to the factorizable emission diagrams, as
illustrated by Figs.~2(a)-2(d), were calculated years ago
in the QCD factorization approach\cite{bbns99,bn651,npb675}. The difference
between the cases that do or do not consider the parton transverse momentum
$k_T$ are very small and can be neglected\cite{nlo05}. The
NLO vertex corrections will be included by  adding a vertex function $V_i(M)$
to the corresponding Wilson coefficients $a_i(\mu)$ \cite{bbns99,npb675},
\beq
a_{1,2}(\mu) &\to& a_{1,2}(\mu)
+\frac{\alpha_s(\mu)}{4\pi}C_F\frac{C_{1,2}(\mu)}{3} V_{1,2}(M) \;, \non
a_i(\mu)&\to &
a_i(\mu)+\frac{\alpha_s(\mu)}{4\pi}C_F\frac{C_{i+1}(\mu)}{3} V_i(M),
\ \  for \ \ i=3,5,7,9, \non a_i(\mu)&\to &
a_i(\mu)+\frac{\alpha_s(\mu)}{4\pi}C_F\frac{C_{i-1}(\mu)}{3} V_i(M),
\ \  for \ \ i=4,6,8,10, \label{eq:aimu-2}
\eeq
where M is the meson emitted from the weak vertex. When $M$ is a
pseudoscalar meson, the vertex functions $V_i(M)$ can be written as
\cite{nlo05,npb675}:
\beq
V_i(M)&=&\left\{ \begin{array}{ll}
12\ln\frac{m_b}{\mu}-18+\frac{2\sqrt{6}}{f_M}\int_{0}^{1}dx\phi_M^A(x)g(x),
& {\rm for}\quad i= 1-4,9,10,\\
-12\ln\frac{m_b}{\mu}+6-\frac{2\sqrt{6}}{f_M}\int_{0}^{1}dx\phi_M^A(x)g(1-x),
&{\rm for}\quad i= 5,7,\\
-6+\frac{2\sqrt{6}}{f_M}\int_{0}^{1}dx\phi_M^P(x)h(x),
&{\rm for} \quad i= 6,8,\\
\end{array}\right.
\label{eq:vim}
\eeq
where $f_M$ is the decay constant of the meson M, and the
hard-scattering functions $g(x)$ and $h(x)$ can be found in Ref.~\cite{zjx2008}.

\item[(b)]{{\bf Quark loops(QL):}}

The contribution from the so-called ``quark loops" is a kind of
penguin correction with the insertion of the four quark operators, as
illustrated by Figs.~3(e) and 3(f). We here include
the quark-loop amplitudes from the operators $O_{1,2}$ and $O_{3-6}$ only.
The quark loops from $O_{7-10}$ will be neglected due to their smallness.

For the $\bar{b}\to \bar{s}$ transition, the contributions from the
various quark loops are given by:
\beq
H_{eff}^{(ql)}&=&-\sum\limits_{q=u,c,t}\sum\limits_{q{\prime}}\frac{G_F}{\sqrt{2}}
V_{qb}^{*}V_{qs}\frac{\alpha_s(\mu)}{2\pi}C^{q}(\mu,l^2)\left(\bar{b}\gamma_\rho
\left(1-\gamma_5\right)T^as\right)\left(\bar{q}^{\prime}\gamma^\rho
T^a q^{\prime}\right),
\eeq
where $l^2$ is  the invariant mass of the gluon, which attaches the quark loops
in Figs.~3(e) and 3(f). The functions $C^{q}(\mu,l^2)$ can be found in Ref.~\cite{zjx2008}.
The ``quark-loop" contribution to the considered $B \to K \etap$ decays can be
written as\cite{zjx2008}
\beq
M_{K\eta}^{(ql)} &=&  <K\eta|{\cal H}_{eff}^{ql}|B>
= \frac{G_F}{\sqrt{2}}\; \sum_{q=u,c,t} V_{qb}^{*}V_{qs}\; \left [ M^{(q)}_{K\eta_s}\; F_2(\phi)
+ M^{(q)}_{\eta_q K}\; F_1(\phi) \right],
\label{eq:mketa}\\
M_{K\etar}^{(ql)} &=&  <K\etar|{\cal H}_{eff}^{(ql)}|B> =
\frac{G_F}{\sqrt{2}}\; \sum_{q=u,c,t} V_{qb}^{*}V_{qs} \; \left [
M^{(q)}_{K\eta_s}\; F_2^\prime(\phi) + M^{(q)}_{\eta_q K}\;
F_1^\prime(\phi) \right], \label{eq:mketap}
\eeq
where  $F_{1,2}^{(\prime)}(\phi)$ are the mixing parameters ( which 
have been defined in previous sections), while the decay amplitudes
$M^{(q)}_{K\eta_s}$ and $M^{(q)}_{\eta_q K}$ are of the form \cite{zjx2008}
\beq
M^{(q)}_{K\eta_s}&=& \frac{8}{\sqrt{6}} C_F^2 m_B^4\;  \int_{0}^{1}d x_{1}d x_{2}\,d
x_{3}\,\int_{0}^{\infty} b_1d b_1 b_3 d b_3\,\phi_B(x_1,b_1)\non &&
\cdot \left\{\left[\left(1+x_3\right) \pka(x_3)\pesa(x_2)
+\rk\left(1-2x_3\right)\left(\pkp(x_3)+\pkt(x_3)\right)\pesa(x_2)
\right.\right.\non && \left.\left. +2\res\pka(x_3)\pesp(x_2)
+2\rk\res\left(\left(2+x_3\right)\pkp(x_3)-
x_3\pkt(x_3)\right)\pesp(x_2)\right] \right.\non && \left. \ \ \cdot
E^{(q)}(t_q,l^2) h_e(x_1,x_3,b_1,b_3) \right.\non && \left. + \left[
2\rk\pkp(x_3)\pesa(x_2) +4\rk\res\pkp(x_3)\pesp(x_2)\right]
\right.\non && \left. \ \ \cdot E^{(q)}(t^{\prime}_q,l^{\prime 2 })
h_e(x_3,x_1,b_3,b_1)\right\}, \label{eq:mqketa}
\eeq
for the $B \to K $ transition, and
\beq
M^{(q)}_{\eta_q K}&=& \frac{8}{\sqrt{6}} C_F^2
m_B^4\; \int_{0}^{1}d x_{1}d x_{2}\,d x_{3}\,\int_{0}^{\infty} b_1d
b_1 b_3d b_3\,\phi_B(x_1,b_1)\non
&& \cdot
\left\{\left[\left(1+x_3\right) \peqa(x_3)\pka(x_2) +\re
\left(1-2x_3\right)\left(\peqp(x_3)+\peqt(x_3)\right)\pka(x_2)
\right.\right.\non && \left.\left. +2\rk \peqa(x_3)\pkp(x_2)
+2\re\rk \left(\left(2+x_3\right)\peqp(x_3)- x_3
\peqt(x_3)\right)\pkp(x_2)\right] \right.\non && \left. \ \ \cdot
E^{(q)}(t_q,l^2) h_e(x_1,x_3,b_1,b_3) \right.\non && \left. + \left[
2\re \peqp(x_3)\pka(x_2) +4\re\rk\peqp(x_3)\pkp(x_2)\right]
\right.\non && \left. \ \ \cdot E^{(q)}(t^{\prime}_q,l^{\prime 2})
h_e(x_3,x_1,b_3,b_1)\right\}, \label{eq:mqetak}
\eeq
for the $B \to \eta$ transition. Here $\re =m_0^q /m_B$ and $\res=m_0^s/m_B$.
The expressions for the evolution functions $E^{(q)}(t_q,l^2)$ and
$E^{(q)}(t^{\prime}_q,l^{\prime 2})$, as well as the hard functions
$h_e(x_1,x_3,b_1,b_3)$ and $h_e(x_3,x_1,b_3,b_1)$ can be found in Ref.~\cite{zjx2008}.

\item[(c)]{ {\bf Chromo-magnetic penguins(MP):}}

This is another kind of penguin correction but with the insertion of 
$O_{8g}$. The corresponding effective weak Hamiltonian for the
$\bar{b}\to \bar{s} g$ transition is of the form\cite{nlo05},
\beq
H_{eff}^{cmp} &=&-\frac{G_F}{\sqrt{2}} \frac{g_s}{8\pi^2}m_b\;
V_{tb}^*V_{ts}\; C_{8g}^{eff} \; \bar{b}_i \;\sigma^{\mu\nu}\; (1-\gamma_5)\;
 T^a_{ij}\; G^a_{\mu\nu}\;  s_j, \label{eq:o8g}
\eeq
where the effective Wilson coefficient $C_{8g}^{eff}= C_{8g} + C_5$ \cite{nlo05}.
The total chromomagnetic penguin contribution to the considered
$B \to K \etap$ decays can be written as
\beq
M_{K\eta}^{(cmp)} &=&  <K\eta|{\cal H}_{eff}^{cmp}|B>
= - \frac{G_F}{\sqrt{2}}\;  \lambda_t \; \left [ M^{(g)}_{K\eta_s}\; F_2(\phi)
+ M^{(g)}_{\eta_q K}\; F_1(\phi) \right],\label{eq:mketa-cmp}\\
M_{K\etar}^{(cmp)} &=&  <K\etar|{\cal H}_{eff}^{cmp}|B> = -
\frac{G_F}{\sqrt{2}}\; \lambda_t \; \left [ M^{(g)}_{K\eta_s}\;
F_2^\prime(\phi) + M^{(g)}_{\eta_q K}\; F_1^\prime(\phi) \right],
\label{eq:mketap-cmp}
\eeq
where the mixing parameters $F_{1,2}^{(\prime)}(\phi)$ have been defined in Sec.II.
The decay amplitudes $M^{(g)}_{K\eta_s}$ and $M^{(g)}_{\eta_q K}$ are
obtained by evaluating the Feynman diagrams Figs.~3(g) and 3(h)\cite{zjx2008},
\beq
M^{(g)}_{K\eta_s}&=& -\frac{8}{\sqrt{6}} C_F^2 m_B^6\;
\int_{0}^{1}d x_{1}dx_{2}\,d x_{3}\,\int_{0}^{\infty} b_1db_1
b_2db_2b_3db_3\,\phi_B(x_1,b_1) \non
&& \cdot
\Big\{\Big\{2(-1+x_3)\pesa(x_2)\pka(x_3)+\res
x_2(1+x_3)[-3\pesp(x_2)+\pest(x_2)]\pka(x_3)\non &&
+\rk[(-3+2x_3+x_3^2)\pkp(x_3)+(-1+2x_3-x_3^2)\pkt(x_3)]\pesa(x_2)\non
&& +3\res \rk
[(-1-x_2+x_3+2x_2x_3)\pkp(x_3)+(1-x_2-x_3+2x_2x_3)\pkt(x_3)]\pesp(x_2)\non
&& +\res
\rk[(-1+x_2+x_3-2x_2x_3)\pkp(x_3)+(1+x_2-x_3-2x_2x_3)\pkt(x_3)]\pest(x_2)\Big\}\non
&& \cdot E_g(t_q)h_g(A,B,C,b_1,b_3,b_2,x_3) \non && - [4\rk
\pesa(x_2) \pkp(x_3)+2\res\rk x_2\pkp(x_3)(3\pesp(x_2)-\pest(x_2))]
\non && \cdot
E_g(t_q^{\prime})h_g(A^{\prime},B^{\prime},C^{\prime},b_3,b_1,b_2,x_1)\Big\},
\label{eq:mpp1} \eeq \beq M^{(g)}_{K\eta_q}&=& -\frac{8}{\sqrt{6}}
C_F^2 m_B^6\; \int_{0}^{1}d x_{1}dx_{2}\,d x_{3}\,\int_{0}^{\infty}
b_1db_1 b_2db_2b_3db_3\,\phi_B(x_1,b_1) \non
&& \cdot \Big\{\Big\{2(-1+x_3)\pka(x_2)\peqa(x_3)+\rk
x_2(1+x_3)[-3\pkp(x_2)+\pkt(x_2)]\peqa(x_3)\non &&
+\re[(-3+2x_3+x_3^2)\peqp(x_3)+(-1+2x_3-x_3^2)\peqt(x_3)]\pka(x_2)\non
&& +3\re \rk
[(-1-x_2+x_3+2x_2x_3)\peqp(x_3)+(1-x_2-x_3+2x_2x_3)\peqt(x_3)]\pkp(x_2)\non
&& +\re
\rk[(-1+x_2+x_3-2x_2x_3)\peqp(x_3)+(1+x_2-x_3-2x_2x_3)\peqt(x_3)]\pkt(x_2)\Big\}\non
&& \cdot E_g(t_q)h_g(A,B,C,b_1,b_3,b_2,x_3) \non && - [4\re
\pka(x_2) \peqp(x_3)+2\re\rk x_2\peqp(x_3)(3\pkp(x_2)-\pkt(x_2))]
\non && \cdot
E_g(t_q^{\prime})h_g(A^{\prime},B^{\prime},C^{\prime},b_3,b_1,b_2,x_1)\Big\},
\label{eq:mpp2}
\eeq
where $\re =m_0^q /m_B$, $\res=m_0^s/m_B$, and $\rk =m_0^K /m_B$. The explicit
expressions of the evolution functions $E_g$ and the hard functions $h_g$ in
Eqs.~(\ref{eq:mpp1}) and (\ref{eq:mpp2}) can be easily found in Ref.~\cite{zjx2008}.

\end{enumerate}

\subsection{The form factors at NLO level}

As mentioned in the Introduction, Li et al. derived the $k_{\rm T}$-dependent NLO hard kernel
$H^{(1)}$ for the $B \to \pi$ transition form factor \cite{prd85074004}.
We here extend their results for the $B \to \pi$ form
factors to the ones for $B \to K$ and $B\to (\eta_q,\eta_s)$ transitions,
under the assumption of $SU(3)$ flavor symmetry. As given in Eq.(56) of Ref.~\cite{prd85074004}, the NLO
hard kernel $H^{(1)}$ can be written as
\begin{eqnarray}
H^{(1)}&=& F(x_1,x_3,\mu,\mu_f,\eta,\zeta_1)H^{(0)}\non & = & {
\alpha_s (\mu_f) C_F \over 4 \pi} \bigg [  {21 \over 4} \ln {\mu^2
\over m_B^2}  -\left ( \ln { m_B^2 \over\zeta_1^2 }  +{13 \over
2}\right) \ln {\mu_f^2 \over m_B^2} +{7 \over 16} \ln^2( x_1 x_3)+{1
\over 8}\ln^2 x_1   \non
&&
+{1 \over 4}\ln x_1\ln x_3 +  \left (2
\ln { m_B^2 \over\zeta_1^2 } +{7 \over 8} \ln \eta - {1 \over 4}
\right) \ln x_1 +  \left ({7 \over 8} \ln \eta - {3 \over 2} \right)
\ln x_3  \non
&& + \left ({15 \over 4} -{7 \over 16} \ln \eta \right ) \ln \eta
 -{1 \over 2} \ln { m_B^2 \over\zeta_1^2 }   \left ( 3
\ln { m_B^2 \over\zeta_1^2 } + 2 \right ) +{101 \over 48} \pi^2
+{219 \over 16} \bigg ] H^{(0)}, \label{pht2}
\end{eqnarray}
where the scale $\zeta_1=25 m_B$, and $\eta = 1-(p_1-p_3)^2 / m_B^2$
is the energy fraction carried by the meson that picks up the spectator quark.
For $B \to K \eta_q, K \eta_s$ decays, the large recoil region
corresponds to the energy fraction  $\eta \sim \textit{O}(1)$.
For $B \to K \eta_c$, $\eta(K) \sim (1-r_{\eta_c}^2)$.
$\mu_f$ is the factorization scale, which is set to be the hard scales
\beq
t^{a}=\max(\sqrt{x_3 \eta } \, m_B ,1/b_1,1/b_3), \quad or \quad
t^{b}=\max(\sqrt{x_1 \eta } \, m_B ,1/b_1,1/b_3),\label{tab}
\eeq
corresponding to the largest energy scales in Figs.~\ref{fig:fig1}(a)
and \ref{fig:fig1}(b), respectively. The renormalization scale $\mu$
is defined as \cite{prd85074004}
\begin{eqnarray}
\mu = t_s (\mu_{\rm f})  = \left \{{\rm Exp} \left[ c_1 + \left(\ln
{m_B^2 \over \zeta_1^2}  +{5 \over 4} \right)  \ln{\mu_{\rm f}^2
\over m_B^2 } \right ]  \, x_1^{c_2 } \, x_3^{c_3} \right \}^{2/21}
\, \mu_{\rm f}, \label{ts function}
\end{eqnarray}
with the coefficients
\begin{eqnarray}
c_1 &=& - \left ({15 \over 4} -{7 \over 16} \ln \eta \right ) \ln
\eta + {1 \over 2} \ln { m_B^2 \over\zeta_1^2 }   \left ( 3 \ln {
m_B^2 \over\zeta_1^2 } + 2 \right ) - {101 \over 48} \pi^2 - {219
\over 16} \,,  \non c_2 &=& - \left ( 2 \ln { m_B^2 \over\zeta_1^2 }
+ {7 \over 8} \ln \eta - {1 \over 4} \right )  \,, \non
c_3 &=& -{7
\over 8} \ln \eta + {3 \over 2}. \nonumber
\end{eqnarray}

At the NLO level, the hard kernel function $H$ can then be written as
\beq
H=H^{(0)}(\alpha_s)+ H^{(1)}(\alpha_s^2)=
\left [ 1+F(x_1,x_3,\mu,\mu_f,\eta,\zeta_1) \right ] H^{(0)}(\alpha_s).
\eeq

\subsection{The NLO contributions to $B \to K \eta_c$ decays}

For $B \to \eta_c K$ decays, the NLO contributions include two parts:
(a) the NLO vertex corrections to these decays, which have 
been taken into account in Ref.~\cite{liux10};
and (b) the NLO contributions to the $B \to K$ transition 
form factors, which is the newly known NLO part.

Since the emitted meson is $\eta_c=c\bar{c}$,
the soft and collinear infrared divergences of the four-vertex correction diagrams
will cancel each other. So these vertex corrections can be calculated without
considering the transverse-momentum effects of the quark at the
end-point region, the same way as in the collinear factorization theorem.

The NLO vertex corrections can be included through the redefinition of 
the Wilson coefficients:
\beq
a_2 &\to & a_2 +\frac{\alpha_s}{4\pi}C_F\frac{C_2}{3}
(-18+12\ln\frac{m_b}{\mu}+f_I), \non a_i &\to &
a_i+\frac{\alpha_s}{4\pi}C_F\frac{C_{i+1}}{3}(-18+12\ln\frac{m_b}{\mu}+f_I),
 \ \ for
\ \ i=3, 9, \non a_j &\to &
a_j-\frac{\alpha_s}{4\pi}C_F\frac{C_{j+1}(\mu)}{3}
(-6+12\ln\frac{m_b}{\mu}+f_I), \ \ for \ \ j=5, 7,
\label{eq:aimu-2ec}
\eeq
where the function $f_I$ is of the form
\beq
f_I &=& \frac{2\sqrt{2 N_c}}{f_{\eta_c}}\int_{0}^{1}dx \phi_{\eta_c}^\nu(x)
\Big[\frac{3(1-2x)}{1-x}\ln x+3(\ln (1-z)-i\pi)+\frac{2z(1-x)}{1-z x}\Big],
\eeq
where $z=m_{\eta_c}^2/m_B^2$.

The NLO contributions to 
the $B \to K$ transition form factors can be included for the $B \to K \eta_c$
decay in the same way as for $B \to K \etap$ decays.

\section{Numerical results and Discussions}\label{sec:n-d}

\subsection{Input parameters}

We use the following input parameters \cite{hfag,pdg2012} in the
numerical calculations(all masses and decay constants are in units of
GeV):
\beq 
f_B &=& 0.21, \quad f_K = 0.16,\quad f_{\eta_c} = 0.4874,
\quad m_{\eta}=0.5475, \quad m_{\eta^{\prime}}=0.9578,\non m_{K^0}
&=& 0.498, \quad m_{K^+} = 0.494, \quad m_{0K} = 1.7, \quad M_B =
5.28,\non  m_b &=& 4.8, \quad m_c = 1.5, \quad m_{\eta_c} = 2.98,
\non M_W &=& 80.41, \quad \tau_{B^0} = 1.53 {\rm
ps}, \quad \tau_{B^+} = 1.638 {\rm ps}. \label{eq:para} \eeq

For the CKM quark-mixing matrix, we adopt the Wolfenstein
parametrization as given in Ref.~\cite{hfag,pdg2012},
\beq
V_{ud}&=&0.9748, \quad V_{us}=\lambda = 0.2246, \quad
|V_{ub}|=3.61\times 10^{-3},\non
V_{cd}&=&-0.225, \quad V_{cs}=0.9748, \quad V_{cb}=0.04197, \non |V_{td}|&=& 8.8 \times
10^{-3}, \quad V_{ts}=-0.042, \quad V_{tb}\approx 1.0,
\label{eq:vckm}
\eeq
with the CKM parameters: $\lambda = 0.2246 \pm 0.0011$,~ $ A = 0.832 \pm 0.017$,~$\bar{\rho} = 0.130 \pm
0.018$, and $\bar{\eta} = 0.350 \pm 0.013$.

\subsection{Form factors at LO and NLO level}

We first calculate and present the pQCD predictions for the form factors at zero momentum transfer
for $B \to K \etap$ decays at the LO and NLO levels, respectively.
In the calculation, we consider three different mixing schemes respectively.

In this paper the form factors $F_0^{B \to \eta}(0)$ and $F_0^{B \to \etar}(0)$ are defined as
\beq
F_0^{B \to \eta}(0)&=&\cos\phi  F_0^{B \to \eta_q}(0)_I, \non
F_0^{B \to \etar}(0)&=&\sin(\phi) F_0^{B \to \eta_q}(0)_I,
\label{eq:ff001}
\eeq
in the ordinary FKS $\eta-\etar$ mixing scheme, and
\beq
F_0^{B \to \eta}(0)&=&\left [ \cos\phi  + \sin\theta\sin\theta_i (1-\cos\phi_G)\right] F_0^{B \to \eta_q}(0)_{II}, \non
F_0^{B \to \etar}(0)&=&\left [ \sin\phi - \cos\theta\sin\theta_i (1-\cos\phi_G)\right] F_0^{B \to \eta_q}(0)_{II},
\label{eq:ff002}
\eeq
in the MS-2 and MS-3 mixing schemes.
One should note that the form factor $F_0^{B \to \eta_q}(0)_I$ is different
from $F_0^{B \to \eta_q}(0)_{II}$ since some relevant parameters in the
distribution amplitudes,
such as $\rho_{\eta_q}=2m_q/m_{qq}$, are different in different mixing schemes.
The pQCD predictions for the numerical values of the form factors for three different
mixing schemes are all listed in Table I, and they are
obtained by using the central values of all  input parameters.
For the relevant mixing angles, we take $\phi=39.3^\circ$ in the MS-1 mixing scheme;
while we take $\theta_i=54.7^\circ$, $\theta=-11^\circ$, 
$\phi=\theta + \theta_i=43.7^\circ$ and
$\phi_G=12^\circ$ in both the MS-2 and MS-3 mixing schemes.
The error comes from the uncertainty of $\omega_b=0.40\pm 0.04$ GeV.

\begin{table}[thb]
\begin{center}
\caption{ The LO and NLO pQCD predictions for the form factors of the $B \to K \etap$
decays  for  three different mixing schemes.}
\label{tableff}
\vspace{0.2cm}
\begin{tabular}{l| c ll } \hline \hline
Form factors& ~~~MS~~~& LO & NLO \\
\hline
$F_0^{B \to \eta_q}(0)$&1 &$0.20 $&$0.26 \pm 0.04 $      \\
                       &2,3&$0.28 $&$0.33\pm 0.06 $\\ \hline
$F_0^{B \to K}(0) $    &all &$0.37 $&$0.43^{+0.07}_{-0.05}$
                             \\ \hline \hline
\end{tabular}\end{center}
\end{table}

From the numerical values of the form factors in Table.~\ref{tableff}, we can see that
(a) the form factors are the same for the $\eta$-$\etar$-$G$ mixing scheme
and the $\eta$-$\etar$-$G$-$\eta_c$ mixing scheme, since the $\eta_c$ component in the
$\eta$-$\etar$-$G$-$\eta_c$ mixing scheme does not affect the evaluation of the
form factors $F_0^{B \to \etap}(0)$ and $F_0^{B \to K }(0)$;
and (b) the NLO contributions also provide $\sim 20\%$ enhancements
to the corresponding form factors.

\subsection{$Br(B \to K \etap)$ in the $\eta$-$\etar$ mixing scheme}\label{subs:ee}

In the B rest frame, the branching ratio of a general $B \to M_2M_3$
decay can be written as
\beq
Br(B\to M_2 M_3) &=& \tau_B\;
\frac{1}{16\pi m_B}\; \chi\; \left | \calm(B\to M_2 M_3) \right|^2,
\label{eq:br-pp}
\eeq
where $\tau_B$ is the lifetime of B meson and 
$\chi\approx 1$ is the phase space factor, which is equal to unit when
the masses of final-state light mesons are neglected.

Using the input parameters  and the wave functions as given in previous sections,
it is straightforward to calculate the CP-averaged branching ratios for the four
$B \to K \etap$ decays considered in different mixing schemes. For the case of the
ordinary $\eta-\etar$ mixing scheme, the pQCD predictions are listed
in Table.~\ref{br1}, where the label ``NLO-1"  refers to the pQCD predictions
with the inclusion of  the same set of NLO contributions as in Ref.~\cite{zjx2008}.
The label ``NLO" in Table II means that all currently known NLO contributions
are included, especially the NLO part of the form factor obtained by
evaluating the Feynman diagrams as shown in Fig.~4 \cite{prd85074004}.
For comparison, we also list the corresponding experimental measurements
\cite{hfag} and the theoretical predictions in the pQCD approach \cite{zjx2008}
and in the QCD factorization (QCDF) approach \cite{npb675}.

\begin{table}[thb]
\begin{center}
\caption{ The pQCD predictions for the branching ratios (in units of
$10^{-6}$) in the ordinary $\eta$-$\etar$ mixing scheme with $\phi=39.3^\circ$.
The label ``NLO" means that all currently known NLO contributions are included. }
\label{br1}
\vspace{0.2cm}
\begin{tabular}{l|  l l l | l| l l} \hline \hline
Channel~~~ & LO~~ & NLO-1 & NLO~~ & Data\cite{hfag} &pQCD\cite{zjx2008}& QCDF\cite{npb675}\\
\hline
$B^0 \to K^0 \eta$   &2.12&2.76&$2.62^{+3.6}_{-1.7}$ &$1.23^{+0.27}_{-0.24}$&$2.1^{+2.6}_{-1.5}$&$1.1 ^{+2.4}_{-1.5}$\\
$B^0 \to K^0 \etar$  &27.9&48.3&$57.2^{+23.7}_{-17.0}$&$66.1\pm 3.1$&$50.3^{+16.8}_{-10.6}$&$46.5^{+41.9}_{-22.0}$\\
$B^+ \to K^+\eta$    &3.83&3.78&$4.0^{+3.8}_{-2.2}$ & $2.4^{+0.22}_{-0.21}$&$3.2^{+3.2}_{-1.8}$&$1.9 ^{+3.0}_{-1.9}$\\
$B^+ \to K^+\etar$   &30.3&49.8&$58.7^{+24.0}_{-17.2}$ &$71.1\pm2.6$&$51.0^{+18.0}_{-10.9}$&$49.1 ^{+45.2}_{-23.6}$\\
\hline\hline
\end{tabular}
\end{center} \end{table}

Of course, the NLO pQCD predictions for the CP-averaged branching ratios
still have large theoretical uncertainties. If we take into account
the effects of the uncertainties of the main input parameters, we find that
\beq
Br(\ B^0 \to K^0 \eta) &=& \left [2.62^{+1.22}_{-0.78}(\omega_b)^{+2.49}_{-1.04}(m_s)^{+0.52}_{-0.48}(f_B) ^{+1.37}_{-1.04}(a^{\eta}_2)
\right ]\times 10^{-6} ,\non
Br(\ B^0 \to K^0 \eta^{\prime}) &=& \left [57.2
^{+16.1}_{-11.2}(\omega_b)^{+12.6}_{-7.00}(m_s)^{+11.4}_{-10.4}(f_B)^{+3.49}_{-2.42}(a^{\eta}_2)
\right ]\times 10^{-6}, \non
Br(\ B^+ \to K^+ \eta) &=& \left
[3.97^{+1.67}_{-1.13}(\omega_b)^{+2.97}_{-1.30}(m_s)^{+0.79}_{-0.72}(f_B)^{+1.57}_{-1.20}(a^{\eta}_2) \right] \times 10^{-6} ,\non
Br(\ B^+ \to K^+ \eta^{\prime}) &=& \left
[58.7^{+16.2}_{-11.2}(\omega_b)^{+13.0}_{-7.20}(m_s)^{+11.7}_{-10.6}(f_B)^{+3.17}_{-2.18}(a^{\eta}_2)
\right ] \times 10^{-6}, \label{eq:brp-eta2ep}
\eeq
where the major errors are induced by the uncertainties of $\omega_b=0.4 \pm 0.04$ GeV,
$m_s=0.13\pm 0.03$ GeV, $f_B=0.21\pm 0.02$ GeV, and the Gegenbauer 
moment $a^{\eta}_2=0.44\pm 0.22$
( here $a_2^\eta$ denotes $a_2^{\eta_q}$ or $a_2^{\eta_s}$), respectively.
The total theoretical errors in the NLO pQCD predictions as shown in the fourth column
of Table II are obtained by adding the four individual theoretical errors in quadrature.
From the numerical results as given in Eq.~(\ref{eq:brp-eta2ep}) and
Table II we make the following points.
\begin{enumerate}
\item[(i)]
By comparing the predictions as listed in the ``NLO-1" column and the ``NLO" column,
one can see that the inclusion of the NLO contributions in the form factors can provide
an $18\%$ enhancement to $Br(B \to K \etar )$. The gap between the pQCD predictions
and the measured values therefore becomes effectively marrow, but there is still a
small difference between the central values of the pQCD predictions and the
data.

\item[(ii)]
For $B \to K \eta $ decays, however, the pQCD predictions for their
branching ratios remain basically unchanged after the inclusion of the
NLO part of the form factors. Although the NLO pQCD predictions for
$Br(B \to K \eta )$ are consistent with the data within one standard
deviation, the central values of the NLO pQCD predictions are still
larger than the measured values by almost a factor of 2.

\item[(iii)]
The pQCD predictions as given in the NLO-1 column agree well with those presented
in Ref.~\cite{zjx2008}(i.e. the results as listed in the ``pQCD" column of Table II),
and the small differences are induced by the variations of some
input parameters, such as the CKM matrix elements.

\item[(iv)]
Although the NLO pQCD predictions for $Br(B \to K \eta )$ and $Br(B
\to K \etar )$ are consistent with the data within one standard
deviation, here we cannot provide a good interpretation for the
observed pattern of $Br(B \to k \etap)$ in the FKS $\eta$-$\etar$
mixing scheme.

\end{enumerate}

\subsection{$Br(B \to K \etap)$ in the $\eta$-$\etar$-$G$ mixing scheme}\label{subs:eeg}

In the $\eta$-$\etar$-$G$ mixing scheme, 
by using the input parameters  and the wave functions as given in previous
sections and fixing the mixing parameters
$\theta=-11^\circ,\phi=43.7^\circ$ and $\phi_G=12^\circ$,
we find the LO and NLO pQCD predictions
for the CP-averaged branching ratios as listed in Table III.
As a comparison, we also show the measured values and the QCDF predictions as given in
Ref.~\cite{npb675} in the last two columns of the Table III.

\begin{table}[thb]
\begin{center}
\caption{ The same as in Table I but for the case of the $\eta$-$\etar$-$G$ mixing
scheme with the mixing parameters $\phi=43.7^\circ, \theta=-11^\circ,
\theta_i=54.7^\circ$, and $\phi_G=12^\circ$. }
\label{br2}
\vspace{0.2cm}
\begin{tabular}{l| l l l| l l} \hline \hline
 Channel~~ & LO~~~~& NLO-1 &NLO~~~& Data\cite{hfag}~~~ & QCDF\cite{npb675}\\
\hline
$B^0 \to K^0 \eta$  &0.90 &1.15 &$1.13^{+2.0}_{-1.0}$ &$1.23^{+0.27}_{-0.24}$ &$1.1 ^{+2.4}_{-1.5}$\\
 $B^0 \to K^0 \etar$&35.2&57.4&$66.5^{+25.9}_{-15.4}$&$66.1\pm 3.1$    &$46.5^{+41.9}_{-22.0}$\\
 $B^+ \to K^+\eta$  &1.98 &2.10 &$2.36^{+2.6}_{-1.5}$ &$2.36^{+0.22}_{-0.21}$ &$1.9 ^{+3.0}_{-1.9}$\\
 $B^+ \to K^+\etar$ &38.9&58.3&$67.3^{+26.0}_{-19.4}$&$71.1\pm2.6$      &$49.1 ^{+45.2}_{-23.6}$\\
\hline\hline
\end{tabular}
\end{center} \end{table}

The NLO pQCD predictions with the inclusion of the major theoretical 
errors are the following:
\beq
Br(\ B^0 \to K^0 \eta) &=& \left [1.13
^{+0.50}_{-0.32}(\omega_b)^{+1.60}_{-0.67}(m_s)^{+0.23}_{-0.20}(f_B)^{+0.99}_{-0.69}(a^{\eta}_2) \right ]\times 10^{-6} ,\non
Br(\ B^0 \to K^0 \eta^{\prime}) &=& \left [66.5
^{+19.6}_{-13.6}(\omega_b)^{+9.9}_{-6.4}(m_s)^{+13.3}_{-12.1}(f_B)^{+3.0}_{-2.1}(a^{\eta}_2)\right ]\times 10^{-6} ,\non
 Br(\ B^+ \to K^+ \eta) &=& \left
[2.36^{+0.97}_{-0.61}(\omega_b)^{+2.05}_{-0.94}(m_s)^{+0.47}_{-0.43}(f_B)^{+1.25}_{-0.90}(a^{\eta}_2)\right
] \times 10^{-6} ,\non
 Br(\ B^+ \to K^+ \eta^{\prime}) &=& \left
[67.3^{+19.6}_{-13.5}(\omega_b)^{+10.2}_{-6.4}(m_s)^{+13.4}_{-12.2}(f_B)^{+2.7}_{-1.8}(a^{\eta}_2)\right ] \times 10^{-6}.
\label{eq:brp-eta2epg}
\eeq
Analogous to the case of MS-1, the major theoretical errors in the 
MS-2 mixing scheme are still induced by the
uncertainties of the input parameters: $\omega_b=0.4 \pm 0.04$ GeV,
$m_s=0.11\pm 0.02$ GeV, $f_B=0.21\pm 0.02$GeV, and Gegenbauer moment
$a^{\eta}_2=0.44\pm 0.22$. The total theoretical errors of the NLO pQCD predictions
in the fourth column of Table III are obtained by adding the four individual
theoretical errors in quadrature.
One can make the following points from the numerical results 
in Eq.~(\ref{eq:brp-eta2epg}) and Table III that
\begin{enumerate}
\item[(i)]
In the  $\eta$-$\etar$-$G$ mixing scheme, the glueball part plays an important
role in improving
the agreement between the pQCD predictions and the data. Even at the leading order,
the pQCD predictions for $Br(B \to K \etar)$ ($Br(B\to K \eta)$) become larger
(smaller) than those in the case of MS-1. The corresponding changes are what
we need  to interpret the data.

\item[(ii)]
For the $B^0 \to K^0\etar$ ($B^+ \to K^+\etar$) decay, the NLO contribution
provides a $ 89\%$ ($73\%$) enhancement to its branching ratio with respect to the LO
prediction. The NLO part of the form factors along provide a $23\%$ enhancement to
the corresponding branching ratios.
The NLO pQCD predictions for both $Br(B^0 \to K^0 \etar)$ and
$Br(B^+ \to K^+ \etar)$ are now in full agreement with the data.

\item[(iii)]
For both $B^0 \to K^0\eta$ and $B^+ \to K^+\eta$ decays, the NLO enhancements
are small in size, and the agreement between the pQCD predictions and the data
also improved effectively due to the inclusion of all of the known NLO contributions.

\item[(iv)]
For all four $B \to K \etap$ decays considered, the differences between the
numerical values as listed in the NLO-1 column and the NLO column in Table III show
the effects of the inclusion of the NLO part of the form factors.
It is easy to see that the NLO pQCD predictions for the branching ratios
are in perfect agreement with the
measured values due to the contribution from the glueball component in the
$\eta$-$\etar$-$G$ mixing scheme and the inclusion of the NLO contributions.

\end{enumerate}

\subsection{$Br(B \to K \etap)$ in the ``$\eta$-$\etar$-$G$-$\eta_c$" mixing scheme}
\label{subs:eegec}

In the ``$\eta$-$\etar$-$G$-$\eta_c$" mixing scheme, by using the input parameters
and the wave functions as given in previous sections and fixing the mixing parameters
$\theta=-11^\circ, \phi=43.7^\circ$, $\phi_G=12^\circ$ and $\phi_Q=11^\circ$,
we find the LO and NLO pQCD predictions
for the CP-averaged branching ratios, which are listed in Table IV.

\begin{table}[thb]
\begin{center}
\caption{ The same as in Table I but for the case of
the ``$\eta$-$\etar$-$G$-$\eta_c$" mixing scheme.
The label NLO means that all known NLO contributions are included. }
\label{br3}
\vspace{0.2cm}
\begin{tabular}{l|  lll|ll} \hline \hline
 Channel~~& LO~~& NLO-1 &~NLO~~ & Data\cite{hfag} & QCDF\cite{npb675}\\
\hline
$B^0 \to K^0 \eta$  &0.67 &0.87 &$0.82^{+1.8}_{-0.8}$ &$1.23^{+0.27}_{-0.24}$ &$1.1 ^{+2.4}_{-1.5}$\\
$B^0 \to K^0 \etar$ &43.5&55.6&$64.8^{+26.8}_{-20.4}$&$66.1\pm 3.1$    &$46.5^{+41.9}_{-22.0}$\\
$B^+ \to K^+\eta$   &1.50 &2.00 &$2.19^{+2.5}_{-1.4}$ &$2.36^{+0.22}_{-0.21}$ &$1.9 ^{+3.0}_{-1.9}$\\
$B^+ \to K^+\etar$  &51.7&56.2&$65.2^{+27.0}_{-20.0}$&$71.1\pm2.6$      &$49.1 ^{+45.2}_{-23.6}$\\
\hline\hline
\end{tabular}
\end{center} \end{table}

In the ``$\eta$-$\etar$-$G$-$\eta_c$" mixing scheme, the contributions from the decay
chain $B \to K \eta_c \to K \etap$ are included.
The NLO pQCD predictions for the CP-averaged branching ratios with the major
theoretical errors are of the form
\beq
Br(\ B^0 \to K^0 \eta) &=& \left [0.82
^{+0.29}_{-0.18}(\omega_b)^{+1.47}_{-0.57}(m_s)^{+0.17}_{-0.14}(f_B)^{+0.92}_{-0.55}(a^{\eta}_2)
 \right ]\times 10^{-6} \label{eq:br0-eta3epgc} ,\non
Br(\ B^0 \to K^0 \eta^{\prime}) &=& \left [64.8
^{+21.2}_{-14.7}(\omega_b)^{+10.3}_{-6.5}(m_s)^{+12.9}_{-11.8}(f_B) \right ]\times
10^{-6} \label{eq:br0-eta4epgc},\non
Br(\ B^+ \to K^+ \eta) &=& \left
[2.19^{+0.75}_{-0.54}(\omega_b)^{+2.06}_{-0.95}(m_s)^{+0.44}_{-0.40}(f_B)^{+1.19}_{-0.86}(a^{\eta}_2)\right
] \times 10^{-6} \label{eq:brp-eta1epgc},\non
Br(\ B^+ \to K^+ \eta^{\prime}) &=& \left
[65.2^{+21.2}_{-14.7}(\omega_b)^{+10.6}_{-6.8}(m_s) ^{+13.0}_{-11.8}(f_B)  \right ]
\times 10^{-6}. \label{eq:brp-eta2epgc}
\eeq
Analogous to the cases of the $\eta$-$\etar$ and $\eta$-$\etar$-$G$  mixing schemes,
the major errors here are also induced by the 
uncertainties of $\omega_b$, $m_s$, $f_B$,  and the 
Gegenbauer coefficient $a^{\eta}_2$, respectively.
For $B \to K \etar$ decays, however, the error induced by the uncertainty
of $a^{\eta}_2=0.44 \pm 0.22$ is very small and has been neglected.
The total theoretical errors of the NLO pQCD predictions in the fourth column
of Table IV are obtained by adding the individual theoretical errors in quadrature.

From Table IV, one can see that the NLO pQCD predictions for $Br(B \to K \etar)$ also agree well with the data.
For $B \to K \eta$ decays, although the central values of the NLO pQCD predictions for
$Br(B \to K \eta)$ are a little smaller than the measured values, but they are
still consistent with the data
within one standard deviation. Since the values of the relevant mixing parameters
$F_C(\theta,\phi_G,\phi_Q)$
and $F_C^\prime(\theta,\phi_G,\phi_Q)$ as defined in Eq.~(\ref{eq:f1f2phip})
are all very small,
\beq
F_C(\theta,\phi_G,\phi_Q)=0.0076, \quad F_C^\prime(\theta,\phi_G,\phi_Q)=0.039,
\eeq
for $(\theta,\phi_G,\phi_Q)=(-11^\circ, 12^\circ,11^\circ)$, the $\eta_c$ contributions
to the $B \to K \etap$ decays are indeed very small.

\subsection{CP-violating asymmetries}

Now we turn to the evaluations of the CP-violating asymmetries of $B
\to K \etap$ decays in the pQCD approach. For $B^+ \to K^+
\etap$ decays, the direct CP-violating asymmetries $\acp$ can be
defined as:
 \beq
\acp^{dir} = \frac{\Gamma(\bar{B}^0 \to \bar{f}) - \Gamma(B^0 \to
f)}{\Gamma(\bar{B}^0 \to \bar{f}) + \Gamma(B^0 \to f)} =
\frac{|\overline{\cal M}_{\bar{f}}|^2 - |{\cal M}_f|^2}{
 |\overline{\cal M}_{\bar{f}}|^2+|{\cal M}_f|^2},
\label{eq:acp1}
\eeq

Using the input parameters and the wave functions as given in previous sections,
it is easy to calculate the direct
CP-violating asymmetries for the considered decays, which are listed in
Table ~\ref{tablecp1}. The major theoretical errors as given in Table ~\ref{tablecp1}
are induced by the uncertainties of the input parameters of $\omega_b$, $m_s$,
and $a_2^\eta$. As a comparison, we also list currently
available data \cite{hfag} and the corresponding QCDF predictions
\cite{npb675}. The label ``NLO" means that all known NLO contributions are
taken into account. For $B^\pm \to K^\pm \eta$ decays, there is a large
direct CP asymmetry ($\acp^{dir}$ ), due to the destructive interference between the
penguin amplitude and the tree amplitude.

\begin{table}[thb]
\begin{center}
\caption{ The LO and NLO pQCD predictions for the direct CP asymmetries
$\acp^{\rm dir}(B^\pm \to K^\pm \eta)$ and
$\acp^{\rm dir}(B^\pm \to K^\pm \etar)$ in the three different mixing
schemes (in units of $10^{-2}$). }
\label{tablecp1} \vspace{0.2cm}
\begin{tabular}{l |c| l l |l l} \hline \hline
 Mode& MS & LO &  NLO & Data\cite{hfag} & QCDF\cite{npb675}\\
\hline
                         &1 &$10.0  $ &$-25.2^{+2.7}_{-1.8}(\omega_b)^{+10.9}_{-12.3}(m_s)^{+8.0}_{-12.2}(a_2^\eta)$&  &\\
$\acp^{\rm dir}(K^\pm \eta) $&2 &$31.1 $ &$-22.9^{+7.6}_{-5.2}(\omega_b)^{+11.6}_{-16.7}(m_s)^{+6.3}_{-7.6}(a_2^\eta)$&$-37\pm 8  $&$-18.9^{+29.0}_{-30.0}$\\
                         &3 &$42.4$  &$-2.8^{+8.5}_{-8.5}(\omega_b)^{+1.9}_{-3.3}(m_s)^{+8.4}_{-3.3}(a_2^\eta)$&\\\hline
                         &1 &$-10.4$ &$-4.4^{+0.7}_{-0.6}(\omega_b)^{+1.1}_{-0.8}(m_s)^{+1.5}_{-1.3}(a_2^\eta)$&  &\\
$\acp^{\rm dir}(K^\pm \etar) $&2 & $-12.2$ &$-5.5^{+0.8}_{-0.8}(\omega_b)^{+0.9}_{-0.7}(m_s)^{+1.5}_{-1.5}(a_2^\eta)$&$1.3^{+1.6}_{-1.7} $&$-9.0^{+10.6}_{-16.2}$\\
                          &3 & $-9.0$  &$-2.3^{+1.1}_{-1.1}(\omega_b)^{+0.4}_{-0.4}(m_s)^{+1.4}_{-1.6}(a_2^\eta)$&\\
\hline \hline
\end{tabular}\end{center}
\end{table}

From the pQCD predictions and the relevant data as listed in Table V, one 
can see the following.
\begin{enumerate}
\item[(i)]
For $B^\pm \to K^\pm \eta$ decays, the LO pQCD predictions for $\acp^{\rm dir}$
in all three mixing schemes  have
a sign opposite that of the measured value. The inclusion of the NLO contributions
changes the sign of the pQCD prediction for $\acp^{\rm dir}$, 
and the NLO pQCD predictions for
$\acp^{\rm dir}(B^\pm \to K^\pm \eta)$ in the cases of MS-1 and MS-2 become
consistent with the data within one standard deviation.
In the ``$\eta$-$\etar$-$G$" mixing scheme, for example, the NLO pQCD predictions
are of the form
\beq
\acp^{dir}(B^\pm \to K^\pm \eta)&=& \left (-22.9^{+15.2}_{-19.1}\right)\times 10^{-2}, \non
\acp^{mix}(B^\pm \to K^\pm \etar)&=& \left ( -5.5^{+1.9}_{-1.8}\right )\times 10^{-2},
\eeq
where the individual errors as shown in Table V are added in quadrature.

\item[(ii)]
In the case of MS-3, however,  the pQCD prediction for $\acp^{dir}(B^\pm \to K^\pm \eta)$
changes its sign after the inclusion of the NLO contributions. The NLO pQCD prediction
is of the form
\beq
\acp^{dir}(B^\pm \to K^\pm \etar)&=& \left ( -2.8^{+10.7}_{-9.7} \right )\times 10^{-2},
\eeq
which is still much smaller in magnitude than the measured 
value in magnitude. There is a clear
difference between the pQCD prediction 
and the data for $\acp^{dir}(B^\pm \to K^\pm \etar)$ in the
 ``$\eta$-$\etar$-$G$-$\eta_c$" mixing scheme.

\item[(iii)]
For $B^\pm \to K^\pm \etar$ decays, the measured value of
$\acp^{\rm dir}(K^\pm \etar)=1.3^{+1.6}_{-1.7}\times 10^{-2} $  is consistent with zero.
The NLO pQCD predictions in the three different mixing schemes agree well with the
data within one  standard deviation, while the consistency between
the pQCD predictions and the data are effectively improved by
the inclusion of the NLO contributions.

\end{enumerate}

As for the CP-violating asymmetries for the neutral decays $B^0 \to
K^0 \etap$, the effects of $B^0-\bar{B}^0$ mixing should be
considered. The CP-violating asymmetry of $B^0(\bar B^0) \to K^0
\etap$ decays are time dependent and can be defined as
\beq
A_{CP} &\equiv& \frac{\Gamma_{\overline{B}_d^0 \to f}(\Delta t) -
\Gamma_{B_d^0 \to f}(\Delta t)}{ \Gamma_{\overline{B}_d^0 \to
f}(\Delta t) + \Gamma_{B_d^0 \to f}(\Delta t)}= {\cal
C}_f\cos(\Delta m \Delta t) + {\cal S}_f\sin (\Delta m \Delta t),
\label{eq:acp-def}
\eeq
where $\Delta m$ is the mass difference between the two $B_d^0$ 
mass eigenstates, and $\Delta t =t_{CP}-t_{tag}
$ is the time difference between the tagged $B^0$ ($\overline{B}^0$)
and the accompanying $\overline{B}^0$ ($B^0$) with opposite b flavor
decaying to the final CP-eigenstate $f_{CP}$ at the time $t_{CP}$.
The direct and mixing-induced CP-violating asymmetries ${\cal C}_f$
( or ${\cal A}_f$ as used by the Belle Collaboration) and ${\cal S}_f$
can be written as
\beq
\acp^{dir}= {\cal C}_f \equiv \frac{ \left |
\lambda\right |^2-1 } {1+|\lambda|^2}, \qquad \acp^{mix}={\cal S}_f
\equiv \frac{ 2 Im (\lambda)}{1+|\lambda|^2}, \label{eq:acp-dm}
\eeq
with the CP-violating parameter $\lambda$
\beq \lambda \equiv \left
( \frac{q}{p} \right )_d \cdot \frac{ \langle f
|H_{eff}|\overline{B}^0\rangle} {\langle f |H_{eff}|B^0\rangle}.
\label{eq:lambda2}
\eeq
By integrating the time variable $t$, one finds the total CP
asymmetries for $B^0 \to K^0 \etap$ decays,
\beq
\acp^{tot}=\frac{1}{1+x^2} A_{CP}^{dir} + \frac{x}{1+x^2} A_{CP}^{mix},
\eeq
where $x=\Delta m/\Gamma=0.774$ \cite{pdg2012}.

\begin{table}[thb]
\begin{center}
\caption{ The pQCD predictions for the direct-, mixing-induced and
total CP asymmetries (in units of $10^{-2}$) for $B^0 \to K^0 \etap$
decays in three different mixing schemes,
and the world averages as given by HFAG \cite{hfag}.}
\label{cp2} \vspace{0.2cm}
\begin{tabular}{l |c| l l | c} \hline \hline
 Mode& ~MS~ & ~~~LO~~~ &  ~~NLO~~ & ~~Data~~ \\\hline
                              &  1  &$ -4.6$
&$-11.1^{+0.7}_{-0.7}(\omega_b)^{+2.9}_{-2.0}(m_s)^{+2.6}_{-3.3}(a_2^\eta)$&$-$  \\
$\acp^{dir}(B^0\to K^0_S \eta)$& 2  &$ -6.6$
&$-16.0^{+0.9}_{-0.8}(\omega_b)^{+5.8}_{-7.8}(m_s)^{+5.3}_{-11.9}(a_2^\eta)$&$-$     \\
                               & 3 &$ -7.8$
&$-19.4^{+0.8}_{-0.0}(\omega_b)^{+8.1}_{-16.2}(m_s)^{+7.4}_{-16.7}(a_2^\eta)$&$-$     \\ \hline
                               & 1  &$ 69.3$
&$66.3^{+0.5}_{-0.3}(\omega_b)^{+2.4}_{-3.3}(m_s)^{+2.2}_{-3.9}(a_2^\eta)$ &$-$   \\
$\acp^{mix}(B^0\to K^0_S \eta)$& 2  &$ 69.9$
&$66.7^{+1.5}_{-1.5}(\omega_b)^{+2.6}_{-6.5}(m_s)^{+2.2}_{-5.2}(a_2^\eta)$ &$-$    \\
                               & 3 &$ 70.3$
&$69.5^{+2.1}_{-2.0}(\omega_b)^{+0.8}_{-2.8}(m_s)^{+2.6}_{-0.5}(a_2^\eta)$ &$-$     \\ \hline
                               & 1   &$ 30.6$
&$25.2^{+0.3}_{-0.2}(\omega_b)^{+3.0}_{-2.9}(m_s)^{+2.7}_{-4.0}(a_2^\eta)$ &$-$   \\
$\acp^{tot}(B^0\to K^0_S \eta)$& 2   &$ 29.7$
&$22.3^{+0.2}_{-0.0}(\omega_b)^{+4.8}_{-8.1}(m_s)^{+4.4}_{-9.9}(a_2^\eta)$ &$-$    \\
                               & 3   &$ 29.2$
&$21.6^{+1.5}_{-1.0}(\omega_b)^{+5.5}_{-11.5}(m_s)^{+4.9}_{-9.2}(a_2^\eta)$ &$-$
                               \\ \hline
                                &1   &$ 1.1  $
&$3.4^{+0.2}_{-0.2}(\omega_b)^{+0.1}_{-0.2}(m_s)^{+0.1}_{-0.1}(a_2^\eta) $&$-$     \\
$\acp^{dir}(B^0\to K^0_S \etar)$&2   &$ 1.0  $
&$3.3^{+0.1}_{-0.2}(\omega_b)^{+0.1}_{-0.1}(m_s)^{+0.1}_{-0.1}(a_2^\eta)$&$1 \pm 9$     \\
                                & 3   &$ 0.9  $
&$3.5^{+0.1}_{-0.1}(\omega_b)^{+0.1}_{-0.2}(m_s)^{+0.2}_{-0.2}(a_2^\eta)$&$-$     \\ \hline
                                & 1   &$ 70.7 $
&$69.8^{+0.1}_{-0.1}(\omega_b)^{+0.2}_{-0.1}(m_s)^{+0.2}_{-0.2}(a_2^\eta)$&$-$     \\
$\acp^{mix}(B^0\to K^0_S \etar)$& 2   &$ 70.8 $
&$70.0^{+0.1}_{-0.1}(\omega_b)^{+0.1}_{-0.1}(m_s)^{+0.1}_{-0.1}(a_2^\eta)$&$64 \pm 11$     \\
                                & 3   &$ 70.8 $
&$70.5^{+0.1}_{-0.1}(\omega_b)^{+0.1}_{-0.0}(m_s)^{+0.1}_{-0.1}(a_2^\eta)$&$-$     \\  \hline
                                & 1   &$ 34.9 $
&$35.9^{+0.2}_{-0.2}(\omega_b)^{+0.0}_{-0.0}(m_s)^{+0.0}_{-0.0}(a_2^\eta)$&$-$     \\
$\acp^{tot}(B^0\to K^0_S \etar)$& 2   &$ 34.9 $
&$36.0^{+0.1}_{-0.2}(\omega_b)^{+0.0}_{-0.0}(m_s)^{+0.0}_{-0.0}(a_2^\eta)$&$-$     \\
                                & 3   &$ 34.8 $
&$36.3^{+0.1}_{-0.1}(\omega_b)^{+0.1}_{-0.1}(m_s)^{+0.1}_{-0.1}(a_2^\eta)$&$-$     \\
 \hline \hline
\end{tabular}
\end{center} \end{table}

In Table ~\ref{cp2}, we show the LO and NLO pQCD predictions for the
direct, the mixing-induced, and the total CP asymmetries for $B^0 \to K^0_S \etap $
decays in the three different mixing schemes.
Analogous to the NLO pQCD predictions for the branching ratios, the label ``NLO"
here means that the inclusion of all currently known NLO contributions are taken
into account.

From the pQCD predictions and currently available experimental
measurements for the CP-violating asymmetries of $B^0 \to K_S^0 \etap$ decays,
one can see the following.
\begin{enumerate}
\item[(i)]
Unlike the cases for the branching ratios, the pQCD predictions for the
CP-violating asymmetries of the neutral
$B^0 \to K_S^0 \etap$ decays are not sensitive to both the NLO contributions and
the choice of the mixing schemes.

\item[(ii)]
The NLO pQCD predictions for $\acp^{\rm dir}(B^0\to K^0_S \etar)$ and
$\acp^{\rm mix}(B^0\to K^0_S \etar)$ have small theoretical error and
agree very well with the measured values:
\beq
\acp^{\rm dir}(B^0\to K^0_S \etar) &=& (3.3\pm 0.3({\rm theory}))\times 10^{-2}, \non
\acp^{\rm mix}(B^0\to K^0_S \etar) &=& (70.3\pm 0.5({\rm theory}))\times 10^{-2},
\eeq
while the measured values are $(1\pm 9)\%$ and $(64\pm 11)\%$, respectively.

\item[(iii)]
The pQCD predictions of $\acp^{\rm dir}(B^0\to K^0_S \eta)\sim -16\%$ and
$\acp^{\rm mix}(B^0\to K^0_S \eta)\sim 67\%$
will be tested by the LHCb and the forthcoming Super-B experiments.

\end{enumerate}

\subsection{Relative strength of the contributions from different sources }

In the pQCD approach at leading order, we usually have the following general
expectations:
\begin{itemize}
\item[(a)]
The factorizable emission diagrams Figs.~1(a) and 1(b) provide the dominant
contribution to the considered $B\to K \etap$ decays;

\item[(b)]
The nonfactorizable spectator diagrams Figs.~1(c) and 1(d) are
strongly suppressed by both the isospin cancelation and the color suppression
and therefore play a minor role;

\item[(c)]
The annihilation diagrams  Figs.~1(e)-1(h) are generally power suppressed in magnitude,
but may provide a large strong phase to produce large CP-violating asymmetries for some
decay modes.
\end{itemize}

In the pQCD approach at next-to-leading order, as discussed in previous sections,
we have made two assumptions.
\begin{itemize}
\item[(a)]
The currently known NLO contributions to $H^{(1)}(\alpha_s^2)$ $-$ 
such as those  coming from the
Feynman diagrams as shown in Figs.~3 and 4 $-$ are the dominant part of
the full NLO contribution.

\item[(b)]
The still missing parts of the NLO contributions from the spectator and annihilation
diagrams as shown in Fig.~5 are small in size and can be safely neglected.
\end{itemize}

Of course, these two assumptions should be examined 
properly before the analytic calculations for the missing parts are performed.
For this purpose, we take the four $B \to K \etap$ decays as an example, and
try to check about the relative strength of the LO or currently known
NLO contributions coming from different sources.

If the LO contributions ( $\propto {\cal O}(\alpha_s)$)
from the nonfactorizable spectator diagrams Figs.~1(c) and 1(d) 
and the annihilation diagrams Figs.~1(e)-1(h)
are already much smaller in size when compared with those from the factorizable emission diagrams Figs.1(a)-1(b),
it is reasonable for us to assume that the still missing next-to-leading
order $O(\alpha_s^2)$ corrections coming from Figs.~5(a)-5(h) should be smaller
than their counterparts at leading order, and therefore much smaller than those
dominant LO contributions; they can therefore be neglected safely.

In order to check whether these general expectations or assumptions  are correct,
we here will firstly decompose the LO decay amplitude $\calm_{\rm LO}$
into different parts according to the corresponding Feynman diagrams,
and then make numerical evaluations for each part and compare their magnitudes directly.
We try to make a simple and clear numerical comparison between the contributions
from different sources.

For the $B^+ \to K^+ \eta$ decay in the $\eta$-$\etar$ mixing scheme,
for example, the decay amplitude $\calm(B^+ \to K^+ \eta)$ at leading order
as given in Eq.~(\ref{eq:kpeta}) \footnote{In the $\eta$-$\etar$ and $\eta$-$\etar$-$G$
mixing schemes, the last term $\calm(B \to \eta_c K)\cdot F_c(\theta,\phi_G,\phi_Q)$ in
Eq.~(\ref{eq:kpeta}) is absent.}  can be rewritten as a sum of three parts
\beq
\calm_{LO}(B^+ \to K^+ \eta)= \calm^{a+b}(K^+ \eta) + \calm^{c+d}(K^+ \eta) +
\calm^{anni}(K^+ \eta),\label{eq:mi-tot}
\eeq
where the decay amplitude $\calm^{a+b}$ is obtained by evaluating the
dominant emission diagrams Figs.~1(a) and 1(b),  $\calm^{c+d}$ refers to
the LO contribution from the spectator diagrams Figs.~1(c) and 1(d),
while $\calm^{\rm anni}$ denotes the LO contribution from the annihilation diagrams
Figs.~1(e)-1(h).

By using the central values of the input parameters and the relevant wave
functions, we make the numerical calculations step by step and then find 
the numerical results (in units of $10^{-4}$)
\beq
\calm_{LO}(B^+ \to K^+ \eta)&=& \underbrace{-1.76 -i 0.37}_{\calm^{a+b}}
+ \underbrace{0.065 -i 0.14}_{\calm^{c+d}}
+ \underbrace{0.03 +i 0.57}_{\calm^{anni}}\non
&=& -1.67 +i 0.062. \label{eq:mi-n}
\eeq
It is easy to see the following.
\begin{itemize}
\item[(a)]
 $\calm^{a+b}=(-1.76 -i 0.37)\times 10^{-4}$ is indeed large and dominant;

\item[(b)]
$\calm^{c+d}=(0.065 -i 0.14)\times 10^{-4}$: its real and imaginary parts 
are all much smaller than
the corresponding parts of both $\calm^{a+b}$ and $\calm^{\rm anni}$;

\item[(c)]
$\calm^{\rm anni}= (0.03 +i 0.57)\times 10^{-4}$; 
its real part is close to zero, but its
imaginary part is large and interferes destructively with $\calm^{a+b}$.
\end{itemize}

Since the branching ratio of the considered decays
are proportional to the square of the decays amplitude $|\calm|^2$,
as shown by Eq.~(\ref{eq:br-pp}), we can define the relative strength of the
individual contribution from different sources as the ratio $R_{\rm LO}$ and then compare
the numerical results directly:
\beq
R_{\rm LO}(K^+\eta)&=&|\calm^{a+b}|^2 : |\calm^{c+d}|^2:|\calm^{anni}|^2:|\calm_{LO}|^2\non
&=& 3.23:0.02:0.33:2.79.
\label{eq:r11}
\eeq
One can see directly from the above numbers that
the contribution from $\calm^{c+d}$ is
less than $1\%$ and can be safely neglected, while
the contributions from $\calm^{\rm anni}$ is also small in magnitude $-$
around $10\%$ of the dominant contribution from emission diagram [ Figs.~1(a) and 1(b)].
This hierarchy of the contributions from different sources agrees 
very well with the general expectations, as stated in the beginning of this subsection.

Using the same methods, we make the similar decompositions and numerical
calculations for the remaining three decay modes $B^0 \to K^0\etap$ and
$B^+\to K^+\etar$, and find the numerical values of the decay amplitudes
and the relative strength.
We make the calculations in both the $\eta$-$\etar$ and $\eta$-$\etar$-$G$
mixing schemes and show all the numerical results in Table \ref{n01}.
For the case of each mixing scheme we use the same input parameters
as those used in the calculation for the branching ratios in Secs.V C and V D
, respectively.

\begin{table}[thb]
\begin{center}
\caption{The LO pQCD predictions for the numerical values (in unit of $10^{-4}$) of the
individual and total decay amplitudes of $B^0 \to K^{0} \etap$ and
$B^+ \to K^+ \etap$ decays, and in the $\eta$-$\etar$
and $\eta$-$\etar$-$G$ mixing scheme.}
\label{n01}
\vspace{0.2cm}
\begin{tabular}{l| c| c| c|c|c|c } \hline \hline
Decay& MS & $\calm^{a+b}$ & $\calm^{c+d}$& $\calm^{anni}$& $\calm_{LO}$ & $R_{LO}$\\ \hline
$ K^0\eta$ & 1& $-1.30 +i 0.04$  & $0.06 -i 0.11$&$-0.06+i 0.53$&$-1.30 +i 0.47$& $1.69:0.01:0.29:1.91$ \\
           & 2& $-0.80 +i 0.03$  & $0.03 -i 0.07$&$0.01 +i 0.54$&$-0.76 +i 0.51$& $0.63:0.01:0.30:0.83$ \\ \hline
$K^0\etar$ & 1& $3.42 +i 0.03$  & $0.25 -i 0.47$&$-0.07 -i 2.85$&$3.40 -i 3.29$ & $11.7:0.29:8.1:22.4$ \\
           & 2& $4.12 +i 0.03$  & $0.24 -i 0.45$&$-0.03 -i 2.94$&$4.32 -i 3.37$ & $17.0:0.27:8.6:30.0$ \\ \hline
$ K^+\eta$ & 1& $-1.76 -i 0.37$  & $0.07 -i 0.14$ &$0.03 +i 0.57$ &$-1.67 +i 0.06$& $3.23:0.02:0.33:2.79$ \\
           & 2& $-1.18 -i 0.52$  & $0.03 -i 0.09$ &$0.10 +i 0.58$ &$-1.05 -i 0.03$& $1.66:0.01:0.35:1.10$ \\ \hline
$K^+\etar$ & 1& $3.65 -i 0.30$  & $0.26 -i 0.54$  &$-0.41 -i 2.99$&$3.50 -i 3.83 $& $13.4:0.36:9.1:26.9$ \\
           & 2& $4.44 -i 0.49$  & $0.25 -i 0.52$  &$-0.38 -i 3.07$&$4.31 -i 4.08$ & $20.0:0.33:9.6:32.2$ \\
\hline\hline
\end{tabular}
\end{center} \end{table}

From the numerical results as shown in Table \ref{n01}, we find the following points:
\begin{enumerate}
\item[(i)]
For all the four considered decays, the factorizable emission diagrams in Figs.~1(a) and 
1(b) provide the dominant contribution to the branching ratios.
In both the MS-1 and MS-2 mixing schemes, we have
\beq
\frac{|\calm^{c+d}|^2}{|\calm^{a+b}|^2} < 0.01, \quad
\frac{|\calm^{\rm anni}|^2}{|\calm^{a+b}|^2} < 0.5,
\label{eq:ra1}
\eeq
for the two $B \to K \eta$ decays, and
\beq
\frac{|\calm^{c+d}|^2}{|\calm^{a+b}|^2} < 0.03, \quad
\frac{|\calm^{\rm anni}|^2}{|\calm^{a+b}|^2} < 0.7, \label{eq:ra2}
\eeq
for  the two $B \to K \etar$ decays.

\item[(ii)]
For all four of the considered decays, one can see from Eqs.(\ref{eq:ra1}) and 
(\ref{eq:ra2}) that
the contribution from the spectator diagrams Figs.~1(c) and 1(d) 
is very small in size,
\beq
\frac{|\calm^{c+d}|^2}{|\calm^{\rm LO}|^2} < 0.03,
\eeq
and therefore can be neglected safely, which is consistent with the general expectation.
Since the LO part $\calm^{c+d}$ is already negligibly small,
it is reasonable for us to neglect the corresponding higher order NLO contribution $\calm_{NLO}^{c+d}$ from
the corresponding spectator diagrams Figs.5(a)-5(d).

\item[(iii)]
For all the four considered decays, the real parts of $\calm^{\rm anni}$ are very small,
but their imaginary parts are relatively large. This leads to a large strong phase,
which is consistent with the general expectation.

\item[(iv)]
For the two $B \to K \etar$ decays, the large imaginary parts of  $\calm^{\rm anni}$ can
also provide an effective enhancement to their branching ratios.
From this point we understand that although the factorizable emission diagrams
Figs.~1(a) and 1(b) provide the dominant contribution to the considered decays, but the LO
contribution from the annihilation diagrams also provide an essential contribution.
Therefore, the NLO contribution $\calm_{\rm NLO}^{\rm anni}$ from the corresponding annihilation
diagrams as shown in Figs.~5(e)-5(h) may be  comparable with other NLO parts,
and thus the  analytical calculations for these diagrams should be done as
soon as possible.

\end{enumerate}

Now we study the relative strength for all known NLO contributions from different sources
and collect all numerical results in Tables \ref{nlo-mi} and \ref{nlo-mir}.

In Table \ref{nlo-mi} we list the numerical values for individual decay amplitudes.
The decay amplitude $\calm_{\rm NLOWC}$ are obtained by 
evaluating the Figs.~1(a)-1(h) using the NLO
Wilson coefficients $C_i(\mu)$, the NLO RG evolution matrix $U(t,m,\alpha)$ 
and $\alpha_s(\mu)$ at the two-loop level.
The label $\calm_{\rm VC}$ denotes the changes of $\calm_{\rm NLOWC}$ when 
only the NLO vertex corrections are also included.
The label $\calm_{\rm ql}$ ($\calm_{\rm mp}$ ) shows the changes 
of $\calm_{\rm NLOWC}$ when only
the NLO contributions from the quark-loops ( the chromo-magnetic penguin) are included.
The label $\calm_{\rm FF}$ shows the variation of $\calm_{NLOWC}$ when only the
$B\to K$ and $B \to \etap$ transition form factors at NLO level are taken into account.

The label $\calm_{\rm VC+ql+mp}$ and $\calm_{\rm NLO}$ in Table \ref{nlo-mi} 
and $\calm_{\rm NLO1}$ in Table \ref{nlo-mir} are defined by the following summations:
\beq
\calm_{\rm VC+ql+mp}&=& \calm_{\rm VC}+ \calm_{\rm ql}+ \calm_{\rm mp}, \\
\calm_{\rm NLO1}&=& \calm_{\rm NLOWC} + \calm_{\rm VC+ql+mp}, \\
\calm_{\rm NLO}^{tot}&=& \calm_{\rm NLO1} + \calm_{\rm FF}.
\eeq
Here  $\calm_{\rm NLO1}$ is equivalent to the total decay amplitude $\calm$ as defined in Eq.(76) of
Ref.~\cite{zjx2008}, and $\calm_{\rm NLO}$ is the decay amplitude when all currently known NLO contributions
are taken into account.

The ratio $R_{\rm NLO}$ in Table \ref{nlo-mir} is defined as
\beq
R_{NLO} = |\calm_{LO}|^2 : |\calm_{NLO1}|^2:|\calm_{NLO}^{tot}|^2.
\label{eq:r22}
\eeq

\begin{table}[thb]
\begin{center}
\caption{The numerical values (in unit of $10^{-4}$)  of the individual NLO
contributions to the decay amplitudes, coming from different sources
in the $\eta$-$\etar$ and $\eta$-$\etar$-$G$ mixing scheme.}
\label{nlo-mi}
\vspace{0.2cm}
\begin{tabular}{l| c| c|c|c|c|c|c } \hline \hline
Decays& MS &$\calm_{NLOWC}$ &$\calm_{VC}$    &$\calm_{ql}$  &$\calm_{mp}$&$\calm_{VC+ql+mp}$& $\calm_{FF}$\\ \hline
$K^0\eta$  &1&$-1.54 +i 0.48$ & $-0.03-i 0.26$ &$-0.32-i0.44$ &$0.35-i0.27$ &$-0.004 -i 0.97$ & $0.05  -i 0.04$ \\
           &2&$-0.96 +i 0.54$ & $-0.07-i 0.08$ &$-0.25-i0.36$ &$0.22-i0.25$ &$-0.10 -i 0.68$ & $-0.001-i 0.02$ \\ \hline
$K^0\etar$ &1&$4.83 -i 3.94$  & $0.63 -i 1.34$ &$1.29 +i1.59$ &$-1.35+i0.37$&$0.57 +i 0.62$  & $0.50 -i 0.25$ \\
           &2&$5.62 -i 4.03$  & $0.61 -i 1.22$ &$1.40 +i1.73$ &$-1.47+i0.36$&$0.54 +i 0.87$  & $0.47 -i 0.24$ \\ \hline\hline
$K^+\eta$  &1&$-1.65 +i 0.11$ & $-0.05 -i 0.35$&$-0.32-i0.44$ &$0.37-i0.27$ &$0.00 -i 1.06$ & $0.001 -i 0.18$  \\
           &2&$-1.13 +i 0.06$ & $-0.10 -i 0.18$&$-0.25-i0.36$ &$0.26-i0.26$ &$-0.09-i 0.80$ & $-0.04 -i 0.13$ \\ \hline
$K^+\etar$ &1&$4.66 -i 4.34$  & $0.65 -i1.38$  &$1.30+i1.59$  &$-1.38+i0.36$&$0.57 +i 0.57$ & $0.46 -i 0.36$ \\
           &2&$5.39 -i 4.57$  & $0.62 -i1.29$  &$1.41+i1.73$  &$-1.50+i0.35$&$0.53 +i 0.79$  & $0.43 -i 0.35$ \\
\hline\hline
\end{tabular}
\end{center} \end{table}

\begin{table}[thb]
\begin{center}
\caption{The numerical values  of $\calm_{\rm LO}$, $\calm_{\rm NLO1}$, $\calm_{\rm NLO}$
(in units of $10^{-4}$) and the ratio $R_{NLO}$ for $B \to K \etap$ decays in
the $\eta$-$\etar$ and $\eta$-$\etar$-$G$ mixing schemes. }
\label{nlo-mir}
\vspace{0.2cm}
\begin{tabular}{l| c| c|c|c|c } \hline \hline
Decay& MS &$\calm_{\rm LO}$    &$\calm_{\rm NLO1}$  & $\calm_{\rm NLO}^{\rm tot}$&$R_{\rm NLO}$\\ \hline
$K^0\eta$  &1&$-1.30 +i 0.47$  & $-1.54-i 0.49$ &$-1.49-i 0.53$ &$1.91:2.61:2.50$  \\
           &2&$-0.76 +i 0.51$  & $-1.05-i 0.14$ &$-1.05-i 0.15$ &$0.84:1.12:1.13$ \\ \hline
$K^0\etar$ &1&$3.40 -i 3.29$   & $5.41 -i 3.32$ &$5.91 -i 3.57$ &$22.4:40.3:47.7$ \\
           &2&$4.32 -i 3.37$   & $6.16 -i 3.16$ &$6.63 -i 3.40$ &$30.0:47.9:55.5$ \\ \hline
$K^+\eta$  &1&$-1.67 -i 0.06$ & $-1.65-i 0.96$ &$-1.65 -i 1.13$&$2.79:3.64:4.00$  \\
           &2&$-1.05 -i 0.03$ & $-1.22-i 0.74$ &$-1.26-i0.87  $&$1.10:2.04:2.35$ \\ \hline
$K^+\etar$ &1&$3.50 -i 3.83 $ & $5.22 -i 3.77$ &$5.69 -i 4.13$ &$26.9:41.5:49.4$ \\
           &2&$4.31 -i 4.08$  & $5.91 -i 3.78$ &$6.34 -i 4.13$ &$32.2:49.2:57.3$ \\
\hline\hline
\end{tabular}
\end{center} \end{table}

From the numerical results as shown in Tables \ref{n01}-\ref{nlo-mir}, 
we find the following points:
\begin{enumerate}
\item[(i)]
For all four $B \to K \etap$ decays,  there are strong cancelations between 
$\calm_{\rm VC}$, $\calm_{\rm ql}$, and $\calm_{\rm mp}$.

\item[(ii)]
For two  $B \to K \eta$ decays,  the corresponding $\calm_{\rm FF}$ 
are also much smaller in magnitude than the
other NLO parts  $\calm_{\rm VC}$, $\calm_{\rm ql}$ and $\calm_{\rm mp}$, 
and also smaller in size than their summation
$\calm_{\rm VC+ql+mp}$.

\item[(iii)]
For the two  $B \to K \etar$ decays,  the corresponding $\calm_{\rm FF}$ 
are  also much smaller than the other
NLO parts  $\calm_{\rm VC}$, $\calm_{\rm ql}$ and $\calm_{\rm mp}$, 
but comparable in size with their summation
$\calm_{\rm VC+ql+mp}$, and therefore all NLO contributions 
together provide the required enhancements to
$Br(B \to K \etar)$ to account for the measured values.

\item[(iv)]
The only missing NLO parts in the pQCD approach  are $\calm_{\rm NLO}^{c+d}$ 
from Figs.~5(a)-5(d) and
$\calm_{\rm NLO}^{anni}$ from Figs.~5(e)-5(h).
They are most probably small in size according to the studies 
in this paper and the general expectations
based on the isospin cancelation and power suppression.

\end{enumerate}

\section{summary }

In this paper, we made a systematic study of the four $B \to K \etap$ decays in the
pQCD factorization approach. We calculated the CP-averaged branching ratios and
CP-violating asymmetries of the four $B \to K\etap$ decays in three different
mixing schemes: the ordinary FKS $\eta$-$\etar$ mixing scheme,
the $\eta$-$\etar$-$G$ mixing scheme, and the $\eta$-$\etar$-$G$-$\eta_c$ mixing scheme.
We considered the full LO contributions and all currently known NLO
contributions to $B \to K \etap$ decays in the pQCD approach.
Besides those NLO contributions considered in Ref.~\cite{zjx2008}, we here took
the newly known NLO part of the $B\to (K, \etap)$ transition form factors 
into account as well.

From our numerical calculations and phenomenological analysis, we find the following points
\begin{enumerate}
\item[(i)]
In all three mixing schemes considered, the NLO pQCD predictions
for the branching ratios and CP-violating asymmetries agree with the
data within one standard deviation, of course, this is partially due to the
still large theoretical errors. However,  the NLO pQCD predictions in the
$\eta$-$\etar$-$G$ mixing scheme provide a nearly perfect
interpretation of the measured values. The NLO pQCD predictions in
MS-2 are the following: 
\beq 
Br(B^0 \to K^0 \eta) &=& (1.13
^{+1.95}_{-1.01}) \times 10^{-6},\non Br(B^0 \to K^0 \etar)&=& (66.5
^{+25.9}_{-19.4}) \times 10^{-6}, \non Br(B^\pm \to K^\pm \eta) &=&
(2.36^{+2.63}_{-1.50}) \times 10^{-6},\non Br(B^\pm \to K^\pm \etar)
&=& (67.3^{+26.0}_{-19.4}) \times 10^{-6}, \label{eq:sbr1} 
\eeq 
for branching ratios, and 
\beq 
\acp^{dir}(B^\pm \to K^\pm \eta)&=&\left
(  -22.9^{+15.2}_{-19.1} \right)\times 10^{-2}, \non
\acp^{dir}(B^\pm \to K^\pm \etar)&=&\left (  -5.5^{+1.9}_{-1.9}
\right)\times 10^{-2}, \non \acp^{dir}(B^0 \to K_S^0 \eta)&=&\left (
-16.0^{+7.9}_{-14.3} \right)\times 10^{-2}, \non \acp^{mix}(B^0 \to
K_S^0 \eta)&=&\left ( 66.7^{+3.7}_{-8.5} \right)\times 10^{-2}, \non
\acp^{dir}(B^0 \to K_S^0 \etar)&=&\left (  3.3\pm 0.3\right)\times
10^{-2},\non \acp^{mix}(B^0 \to K_S^0 \etar)&=&\left (  70.0\pm 0.3
\right)\times 10^{-2}, \label{eq:sacp1} 
\eeq 
for the CP-violating
asymmetries,  where the individual theoretical errors have been
combined in quadrature.

\item[(ii)]
For the $B^0 \to K^0 \etar$ and $B^+ \to K^+ \etar$ decays, the NLO contributions provide significant
enhancements to their branching ratios. In the $\eta$-$\etar$-$G$ mixing scheme,
for example,  the NLO contribution provides a $ 89\%$ ($73\%$) enhancement to
$Br(B^0 \to K^0\etar)$ ($Br(B^+ \to K^+\etar)$ ) with respect to the LO
prediction, such enhancements play a key role in our effort to resolve the
$K\etap$ puzzle and to understand the patten of the $Br(B \to K \etap)$.

\item[(iii)]
For the $B^0 \to K^0 \eta$ and $B^+ \to K^+ \eta$ decays, the inclusion
of the NLO contributions only leads to a relatively small changes to
their branching ratios, but the resulting variations are in the right
direction and helpful for us in improving  the consistency between the
pQCD predictions and the measured values. In the $\eta$-$\etar$-$G$
mixing scheme, for instance, the central values of the pQCD predictions
are $Br(B^0\to K^0\eta)=0.90\times 10^{-6}$ and $Br(B^+\to K^+\eta)=1.98\times
10^{-6}$ at the leading order, these changed to $Br(B^0\to
K^0\eta)=1.13\times 10^{-6}$ and $Br(B^+\to K^+\eta)=2.36\times
10^{-6}$ when the NLO contributions were taken into account, while
the corresponding measured values are $1.23^{+0.27}_{-0.23} \times
10^{-6}$ and $ 2.36^{+0.22}_{-0.21}\times 10^{-6}$ respectively.

\item[(iv)]
By comparing the pQCD predictions as given in the  ``NLO1" and ``NLO" columns 
in Tables II-IV, one can directly see 
the effects of NLO form factors:  the NLO part $\calm_{\rm FF}$ of 
the $B \to K$ and $B \to \etap$ form factors
can produce an about $20\%$ enhancement to the branching ratios $Br(B \to K \etar)$, which
plays an important role in closing the gap between the pQCD predictions and the relevant data.

\item[(v)]
In the $\eta$-$\etar$-$G$-$\eta_c$ mixing scheme, the decay chain 
$B \to K\eta_c\to K \etap$ can
provide an effective enhancement to the branching ratios at the leading-order,
but when the large NLO contributions are taken into account, 
the effects of the $\eta_c$ component
become unimportant.

\item[(vi)]
For $B^\pm \to K^\pm \eta$ decays, the LO pQCD predictions for $\acp^{dir}$ in all three mixing schemes
have a sign opposite that of the measured value.  The inclusion of the NLO 
contributions changed the sign of the pQCD predictions for 
$\acp^{\rm dir}$, while the NLO pQCD predictions for
$\acp^{\rm dir}(B^\pm \to K^\pm \eta)$ in the cases of the 
MS-1 and MS-2 are now becoming consistent with the data
within one standard deviation, but the NLO pQCD prediction 
for $\acp^{\rm dir}(B^\pm \to K^\pm \eta)$ in the MS-3 case 
is still much smaller in magnitude than the measured value.

\item[(vii)]
For $\acp^{dir}(B^\pm \to K^\pm \etar)$, the NLO pQCD predictions agree 
with the data within one standard deviation,
while the consistency between the pQCD predictions and the data is 
improved by the inclusion of the NLO contributions.

\item[(viii)]
For the direct and mixing-induced CP-violating asymmetries
$\acp^{\rm dir}(B^0 \to K_S^0 \etap)$ and  $\acp^{\rm mix}(B^0 \to K_S^0 \etap)$,
the pQCD predictions have a weak dependence on the NLO contributions 
and the choice of different
mixing schemes. For $\acp^{\rm dir,mix}(B^0\to K^0_S \etar)$, for example, 
the NLO pQCD predictions are
$\acp^{\rm dir}(B^0\to K^0_S \etar)\approx 3\%$ 
and $\acp^{\rm mix}(B^0\to K^0_S \etar)\approx 70\%$,
which  are well consistent with the measured values 
of $(1 \pm 9)\%$ and $(64\pm 11)\%$ respectively.

\item[(ix)]
The factorizable emission diagrams Figs.~1(a) and 1(b) provide the dominant 
contribution to the considered decays.
The LO contribution $\calm^{c+d}$ from the spectator diagrams Figs.~1(c) and 1(d) 
is already
less than $3\%$ of the total contribution,
the next-to-leading order contributions $\calm_{NLO}^{c+d}$ from the
Figs.~5(a)-5(d) are the higher-order contributions and therefore should be
smaller than their LO counterpart $\calm^{c+d}$.
Consequently, it is reasonable for us to neglect $\calm_{\rm NLO}^{c+d}$ from
the spectator diagrams Figs.~5(a)-5(d).

\item[(x)]
The real part of $\calm^{\rm anni}$ is always negligibly small, but its imaginary part
is relatively large and leads to a large strong phase,
which can also produce an effective enhancement to the branching ratios
of the considered decays.
Although $|\calm_{\rm NLO}^{\rm anni}|$ is most possibly much smaller than its LO counterpart
$|\calm^{\rm anni}|$, but the still missing NLO contribution $\calm^{\rm anni}_{\rm NLO}$
from Figs.~5(e)-5(h) may be comparable in size
with $\calm_{\rm FF}$, and should be calculated as soon as possible.

\end{enumerate}

\begin{acknowledgments}

The authors are very grateful  to Hsiang-nan Li, Cai-Dian Lu  and Xin Liu for helpful
discussions. This work is supported by the National Natural Science
Foundation of China under the Grant Nos.~10975074 and~11235005;
and by the Project on Graduate Students¡¯Education and Innovation of Jiangsu
Province, under Grant No. CXLX11-0867.

\end{acknowledgments}


\begin{appendix}

\section{Distribution Amplitudes }\label{sec:da}

The expressions for the relevant distribution amplitudes (DAs) of the K
meson are the following \cite{ball98,ball05}:
\begin{eqnarray}
\pka(x) &=&  \frac{f_K}{2\sqrt{2N_c} }  6x (1-x)
    \left[1+a_1^{K}C^{3/2}_1(t)+a^{K}_2C^{3/2}_2(t)+a^{K}_4C^{3/2}_4(t)
  \right],\label{piw1}\\
 \pkp(x) &=&   \frac{f_K}{2\sqrt{2N_c} }
   \left\{ 1+(30\eta_3-\frac{5}{2}\rho^2_{K})C^{1/2}_2(t)
   -3\left[ \eta_3\omega_3+\frac{9}{20}\rho^2_K
   (1+6a^K_2)\right]C^{1/2}_4(t)\right\}, \ \ \\
 \pkt(x) &=&  - \frac{f_K}{2\sqrt{2N_c} } t
   \left[ 1+6(5\eta_3-\frac{1}{2}\eta_3\omega_3-\frac{7}{20}\rho^2_{K}
   -\frac{3}{5}\rho^2_Ka_2^{K})
   (1-10x+10x^2)\right] ,\quad\quad\label{piw}
\end{eqnarray}
with the mass ratio $\rho_K=m_K/m_{0K}$.  The Gegenbauer moments are of the form \cite{ball98}:
\beq
a^K_1=0.2 ,\quad a^K_2=0.25, \quad a^K_4=-0.015.
\eeq
The values of the other parameters are $\eta_3=0.015$
and $\omega=-3.0$. Finally, the Gegenbauer polynomials $C^{\nu}_n(t)$
are given as:
\beq
C^{1/2}_2(t)&=&\frac{1}{2}(3t^2-1), \quad
C^{1/2}_4(t)=\frac{1}{8}(3-30t^2+35t^4), \non C^{3/2}_1(t)&=&3t,
\quad C^{3/2}_2(t)=\frac{3}{2}(5t^2-1),\non
C^{3/2}_4(t)&=&\frac{15}{8}(1-14t^2+21t^4), \label{eq:c124}
\eeq
with $t=2x-1$.

The distribution amplitudes $\phi_{\eta_q}^{A,P,T}$ are given as
\cite{ball98}:
\begin{eqnarray}
 \phi_{\eta_q}^A(x) &=&  \frac{f_q}{2\sqrt{2N_c} }
    6x (1-x)
    \left[1+a^{\eta_q}_1C^{3/2}_1(2x-1)+a^{\eta_q}_2 C^{3/2}_2(2x-1)
    \right.\non && \left.+a^{\eta_q}_4C^{3/2}_4(2x-1)
  \right],\label{piw11}\\
 \phi_{\eta_q }^P(x) &=&   \frac{f_q}{2\sqrt{2N_c} }
   \left[ 1+(30\eta_3-\frac{5}{2}\rho^2_{\eta_q } )C^{1/2}_2(2x-1)
\right.\non && \left.
   -3\left\{\eta_3\omega_3+\frac{9}{20}\rho^2_{\eta_q }
   (1+6a^{\eta_q }_2)\right\} C^{1/2}_4(2x-1)\right]  ,\\
 \phi_{\eta_q}^T(x) &=&  \frac{f_q}{2\sqrt{2N_c} } (1-2x)
   \left[ 1+6\left (5\eta_3-\frac{1}{2}\eta_3\omega_3
   -\frac{7}{20}\rho^2_{\eta_q}-\frac{3}{5}\rho^2_{\eta_q }a_2^{\eta_q}\right )
   \right. \non && \left.
   \cdot \left (1-10x+10x^2 \right )\right] ,\quad\quad\label{piw4}
 \end{eqnarray}
where $\rho_{\eta_q}=2m_q/m_{qq}$,$a^{\eta_q}_1=a_1^\pi=0$,
$a^{\eta_q}_2=a_2^\pi=0.44\pm 0.22$, $a^{\eta_q}_4=a_4^\pi=0.25$,
and the Gegenbauer polynomials $C^{\nu}_n(t)$ have been given in
Eq.~(\ref{eq:c124}). As for the wave function and the corresponding
DAs of the $s\bar{s}$ components, we also use the same form as
$q\bar{q}$ but with some parameters changed: $\rho_{\eta_s}
=2m_s/m_{ss}$, $a^{\eta_s}_i=a^{\eta_q}_i$ for $i=1,2,4$.

\section{ Related Hard Functions}\label{sec:hf}

The hard scales appearing in the decay amplitudes are chosen as
\beq
t_a&=&\mbox{max}\{{\sqrt {x_3}M_{B},1/b_1,1/b_3}\},\non
t_a^\prime&=&\mbox{max}\{{\sqrt {x_1}M_{B},1/b_1,1/b_3}\},\non
t_b&=&\mbox{max}\{\sqrt {x_1x_3}M_{B},\sqrt{|1-x_1-x_2|x_3}M_{B},1/b_1,1/b_2\},\non
t_b^\prime&=&\mbox{max}\{\sqrt{x_1x_3}M_{B},\sqrt {|x_1-x_2|x_3}M_{B},1/b_1,1/b_2\},\non
t_c&=&\mbox{max}\{\sqrt{1-x_3}M_{B},1/b_2,1/b_3\},\non
t_c^\prime &=&\mbox{max}\{\sqrt {x_2}M_{B},1/b_2,1/b_3\},\non
t_d&=&\mbox{max}\{\sqrt {x_2(1-x_3)}M_{B}, \sqrt{1-(1-x_1-x_2)x_3}M_{B},1/b_1,1/b_2\},\non
t_d^\prime&=&\mbox{max}\{\sqrt{x_2(1-x_3)}M_{B},\sqrt{|x_1-x_2|(1-x_3)}M_{B},1/b_1,1/b_2\},\non
t_e&=&\mbox{max}\{{\sqrt  {x_3(1-r_{\eta_c}^2)}M_{B},1/b_1,1/b_3}\},\non
t_e^\prime&=&\mbox{max}\{{\sqrt {x_1(1-r_{\eta_c}^2)}M_{B},1/b_1,1/b_3}\},\non
t_f&=&\mbox{max}\{\sqrt{x_1x_3(1-r_{\eta_c}^2)}M_{B},\sqrt{|(-1+x_1+x_2)[x_3+(1-x_2-x_3)r_{\eta_c}^2]
+r_{\eta_c}^2|}M_{B},\non && 1/b_1,1/b_2\},\non
t_f^\prime&=&\mbox{max}\{\sqrt{x_1x_3(1-r_{\eta_c}^2)}M_{B},
\sqrt{|(x_1-x_2)[x_3+(x_2-x_3)r_{\eta_c}^2]+r_{\eta_c}^2|}M_{B},\non
&& 1/b_1,1/b_2\}.
\eeq

The hard functions $h_i$s appearing in the decay amplitudes are defined by
\beq
h_e(x_1,x_3,b_1,b_3)&=&\left[\theta(b_1-b_3)I_0(\sqrt{x_3}M_{B}b_3)K_0(\sqrt{x_3}
M_{B}b_1)\right. \non
&&
\left.+\theta(b_3-b_1)I_0(\sqrt{x_3}M_{B}b_1)K_0(\sqrt{x_3}M_{B}b_3)\right]K_0(\sqrt{x_1x_3}M_{B}b_1)S_t(x_3),\non
h_n(x_1,x_2,x_3,b_1,b_2)&=&\left[\theta(b_2-b_1)K_0(\sqrt{x_1x_3}M_{B}b_2)I_0(\sqrt{x_1x_3}M_{B}b_1)\right.
\non &&\;\;\;\left.
+\theta(b_1-b_2)K_0(\sqrt{x_1x_3}M_{B}b_1)I_0(\sqrt{x_1x_3}M_{B}b_2)\right]\non
&&\times
\left\{\begin{array}{ll}\frac{i\pi}{2}H_0^{(1)}(\sqrt{(x_2-x_1)x_3}
M_{B}b_2),& x_1-x_2<0\\
K_0(\sqrt{(x_1-x_2)x_3}M_{B}b_2),& x_1-x_2>0
\end{array}  \right. ,
\eeq
\beq
h_a(x_2,x_3,b_2,b_3)&=&(\frac{i\pi}{2})^2
S_t(x_3)\Big[\theta(b_2-b_3)H_0^{(1)}(\sqrt{x_3}M_{B}b_2)J_0(\sqrt
{x_3}M_{B}b_3)\non
&&\;\;+\theta(b_3-b_2)H_0^{(1)}(\sqrt {x_3}M_{B}b_3)J_0(\sqrt
{x_3}M_{B}b_2)\Big]H_0^{(1)}(\sqrt{x_2x_3}M_{B}b_2),\non
h_{na}(x_1,x_2,x_3,b_1,b_2)&=&\frac{i\pi}{2}\left[\theta(b_1-b_2)H^{(1)}_0(\sqrt
{x_2(1-x_3)}M_{B}b_1)J_0(\sqrt {x_2(1-x_3)}M_{B}b_2)\right. \non
&&\;\;\left.
+\theta(b_2-b_1)H^{(1)}_0(\sqrt{x_2(1-x_3)}M_{B}b_2)J_0(\sqrt
{x_2(1-x_3)}M_{B}b_1)\right]\non
&&\;\;\;\times K_0(\sqrt{1-(1-x_1-x_2)x_3}M_{B}b_1),
\eeq
\beq
h_{na}^\prime(x_1,x_2,x_3,b_1,b_2)&=&\frac{i\pi}{2}\left[\theta(b_1-b_2)H^{(1)}_0(\sqrt
{x_2(1-x_3)}M_{B}b_1)J_0(\sqrt{x_2(1-x_3)}M_{B}b_2)\right. \non
&&\;\;\;\left. +\theta(b_2-b_1)H^{(1)}_0(\sqrt
{x_2(1-x_3)}M_{B}b_2)J_0(\sqrt{x_2(1-x_3)}M_{B}b_1)\right]\non
&&\;\;\;\times
\left\{\begin{array}{ll}\frac{i\pi}{2}H^{(1)}_0(\sqrt{(x_2-x_1)(1-x_3)}M_{B}b_1),&
x_1-x_2<0\\
K_0(\sqrt {(x_1-x_2)(1-x_3)}M_{B}b_1),& x_1-x_2>0\end{array}\right.
,
\eeq
where $H_0^{(1)}(z) = \mathrm{J}_0(z) + i\, \mathrm{Y}_0(z)$.
\beq
h_e^\prime(x_1,x_3,b_1,b_3)&=&\left[\theta(b_1-b_3)I_0(\sqrt{\alpha}M_{B}b_3)K_0(\sqrt{\alpha} M_{B}b_1)\right.\non
&&
\left.+\theta(b_3-b_1)I_0(\sqrt{\alpha}M_{B}b_1)K_0(\sqrt{\alpha}M_{B}b_3)\right]K_0(\sqrt{\beta}M_{B}b_1)S_t(x_3),\non
\eeq
where $\alpha = x_3(1-r_{\eta_c}^2)$, ~$\beta=x_1x_3(1-r_{\eta_c}^2)$.
\beq
h_n^\prime(x_1,x_2,x_3,b_1,b_2)&=&\left[\theta(b_2-b_1)K_0(\sqrt{\alpha'}M_{B}b_2)I_0(\sqrt{\alpha'}M_{B}b_1)\right.
\non &&\;\;\;\left.
+\theta(b_1-b_2)K_0(\sqrt{\alpha'}M_{B}b_1)I_0(\sqrt{\alpha'}M_{B}b_2)\right]\non
&&\times
\left\{\begin{array}{ll}\frac{i\pi}{2}H_0^{(1)}(\sqrt{|\beta^{'2}|}
M_{B}b_2),& \beta^{'2}<0\\
K_0(\sqrt{|\beta^{'2}|}M_{B}b_2),& \beta^{'2}>0
\end{array} \right. ,
\eeq
where $\alpha' = x_1x_3(1-r_{\eta_c}^2)$, and $\beta^{'2} =
(x_1-x_2)[x_2r_{\eta_c}^2+x_3(1-r_{\eta_c}^2)]+r_{\eta_c}^2$.

The function $S_t(x)$ has been parametrized as \cite{prd65,hnlprd}
\beq
S_t(x)=\frac{2^{1+2c}\Gamma(3/2+c)}{\sqrt \pi \Gamma(1+c)}[x(1-x)]^c,
\label{eq:stxi}
\eeq
with $c=0.3$.

The evolution factors $E^{(\prime)}_e$, and $E^{(\prime)}_a$,
appearing in the decay amplitudes are given by
\beq
E_e(t)&=&\alpha_s(t) \exp[-S_B(t)-S_3(t)],\non
 E'_e(t)&=&\alpha_s(t)
 \exp[-S_B(t)-S_2(t)-S_3(t)]|_{b_1=b_3},\non
E_a(t)&=&\alpha_s(t)
 \exp[-S_2(t)-S_3(t)],\non
E'_a(t)&=&\alpha_s(t) \exp[-S_B(t)-S_2(t)-S_3(t)]|_{b_2=b_3},
\eeq
where the Sudakov exponents are defined as \cite{kls01,lu2001,li2003}
\beq
S_B(t)&=&s\left(x_1\frac{M_{B}}{\sqrt
2},b_1\right)+\frac{5}{3}\int^t_{1/b_1}\frac{d\bar \mu}{\bar
\mu}\gamma_q(\alpha_s(\bar \mu)),\label{eq:sb}\non
S_2(t)&=&s\left(x_2\frac{M_{B}}{\sqrt
2},b_2\right)+s\left((1-x_2)\frac{M_{B}}{\sqrt
2},b_2\right)+2\int^t_{1/b_2}\frac{d\bar \mu}{\bar
\mu}\gamma_q(\alpha_s(\bar \mu)),
\eeq
with the quark anomalous dimension $\gamma_q=-\alpha_s/\pi$. Replacing the
variables $(x_2,b_2)$ in $S_2$ by $(x_3,b_3)$, we get the expression for
$S_3$. At the one-loop order, the explicit expression of the  function $s(Q,b)$ is
\cite{kls01,lu2001}:
\beq
s(Q,b)&=&~~\frac{A^{(1)}}{2\beta_{1}}\hat{q}\ln\left(\frac{\hat{q}}
{\hat{b}}\right)-
\frac{A^{(1)}}{2\beta_{1}}\left(\hat{q}-\hat{b}\right)+
\frac{A^{(2)}}{4\beta_{1}^{2}}\left(\frac{\hat{q}}{\hat{b}}-1\right)
\non
&&-\left[\frac{A^{(2)}}{4\beta_{1}^{2}}-\frac{A^{(1)}}{4\beta_{1}}
\ln\left(\frac{e^{2\gamma_E-1}}{2}\right)\right]
\ln\left(\frac{\hat{q}}{\hat{b}}\right)+ \cdots
\eeq
where the variables are defined by
\beq
\hat q\equiv \mbox{ln}[Q/(\sqrt 2\Lambda)],~~~ \hat b\equiv
\mbox{ln}[1/(b\Lambda)],
\eeq
and the coefficients $A^{(i)}$ and $\beta_1$ are of the form
\beq
\beta_1=\frac{33-2n_f}{12},\quad A^{(1)}=\frac{4}{3}, \quad A^{(2)}=\frac{67}{9}
-\frac{\pi^2}{3}-\frac{10}{27}n_f+\frac{8}{3}\beta_1\mbox{ln}(\frac{1}{2}e^{\gamma_E}),
\eeq
where $n_f$ is the number of the quark flavors and $\gamma_E$ is the
Euler constant.

\end{appendix}


\end{document}